\documentclass[fleqn,pdflatex,10pt]{wlscirep}
\usepackage{subcaption}
\usepackage[utf8]{inputenc}
\usepackage[T1]{fontenc}

\newcommand{\GF}[1]{\mathit{GF}_{#1}}

\newcommand{\POIattr}[1]{P_{#1}}
\newcommand{\DB}{\mathcal{D}}
\newcommand{\uid}{\mathit{uid}}
\newcommand{\seq}{\mathit{p}}

\title{Data-Driven Discrete Geofence Design Using Binary Quadratic Programming}

\author[1,*]{Keisuke Otaki}
\author[1]{Akihisa Okada}
\author[1]{Tadayoshi Matsumori}
\author[1]{Hiroaki Yoshida}
\affil[1]{Toyota Central R\&D Labs., Inc., Bunkyo-ku, Tokyo, 112-0004, Japan}

\affil[*]{otaki@mosk.tytlabs.co.jp}

\begin{abstract}
Geofences have attracted significant attention in the design of spatial and virtual regions for managing and engaging spatiotemporal events.
By using geofences to monitor human activity across their boundaries, content providers (e.g., travel agencies) can create spatially triggered events that include notifications about points of interest within a geofence by pushing spatial information to the devices of users.
Traditionally, geofences were hand-crafted by providers.
In addition to the hand-crafted approach, recent advances in collecting human mobility data through mobile devices can accelerate the automatic and data-driven design of geofences, also known as the geofence design problem.
Previous approaches assume circular shapes; thus, their flexibility is insufficient, and they can only handle geofence-based applications for large areas with coarse resolutions.
A challenge with using circular geofences in urban and high-resolution areas is that they often overlap and fail to align with political district boundaries and road segments, such as one-way streets and median barriers.
In this study, we address the problem of extracting arbitrary shapes as geofences from human mobility data to mitigate this problem.
In our formulation, we cast the existing optimization problems for circular geofences to 0-1 integer programming problems to represent arbitrary shapes.
Although 0-1 integer programming problems are computationally hard, formulating them as quadratic (unconstrained) binary optimization problems enables efficient approximation of optimal solutions, because this allows the use of specialized quadratic solvers, such as the quantum annealing, and other state-of-the-art algorithms.
We then develop and compare different formulation methods to extract discrete geofences.
Through computational experiments with quantitative comparisons using different parameters, we confirmed that our new modeling approach enables flexible geofence design.
\end{abstract}

\begin{document}

\flushbottom
\maketitle
\thispagestyle{empty}

\section*{Introduction}
\label{sec-intro}

With the widespread adoption of mobile devices such as smartphones and smartwatches, mobility service providers can design diverse experiences and applications, including mobility-on-demand services and flexible food delivery services.
In such applications, providers can leverage the location data of users to effectively deliver engaging information about their merchandise, transitioning users from traditional printed maps, and enabling users' ubiquitous access to multi-dimensional information from applications.
In practice, the progress of navigation applications enables us to access navigation applications easily~\cite{Market2019}.
Conversely, mobility activities have declined and community fragmentation has increased~\cite{Hara2021, Chen2023, Nilforoshan2023}, particularly since the COVID-19 pandemic. 
New mobility services with novel concepts are promising for addressing the above issues by promoting human travel. 
These systems can contribute to various societal benefits, including economic growth, urban development, and enhanced well-being. 
While traditional navigation systems were primarily used for large-scale mobility modes such as vehicles and ships, they have become equally crucial for micro-scale movements in modern applications, including cycling, micro-mobility, and walking~\cite{Speake2015}.
The ability to exchange geospatially contextualized information with users is a fundamental component when discussing the new mobility experiences.

This study focuses on the geofence design problem when designing mobility experiences by leveraging spatial information.
In designing mobility experiences, area-based information has traditionally been designed primarily by service providers and system architects around specific locations (points of interest~(POIs) or landmarks), with a particular emphasis on entry and exit events within these areas.
As diverse datasets continue to increase in quantity, data-driven approaches have become increasingly important in information communication design~\cite{Pappalardo2023, Tong2022}.
Utilizing mobility data such as GPS records, we can effectively enhance the travel experiences of users by exchanging information about specific regions (e.g., an event area in a city), a concept known as \textit{geofence/geofencing}. 
Current technology enables precise detection when users enter or exit geofences defined by service providers, utilizing both GPS location tracking and network connectivity via mobile data and Wi-Fi~\cite{Reclus2009, Namiot2013, Nakagawa2013, Nakagawa2014}.
They can also be applied in tracking~\cite{Thoren2024}, vehicle controls~\cite{Hansen2021, Hansen2024, Volvo2025}, and information notifiers~\cite{Sasaki2024, Amshavalli2025}.
Recent research has also examined the behavioral patterns of users concerning geofences, highlighting it as a significant concept tied to geographic space~\cite{Shevchenko2024}.
Commercially, applications could include directing users to lively areas to promote walking tours, distributing store coupons to boost consumption, or guiding users to avoid congested areas, thus reducing traffic congestion. 
The application of these technologies in designing next-generation experiences is promising.

\begin{figure}[h]
\centering
\includegraphics[width=0.9\linewidth]{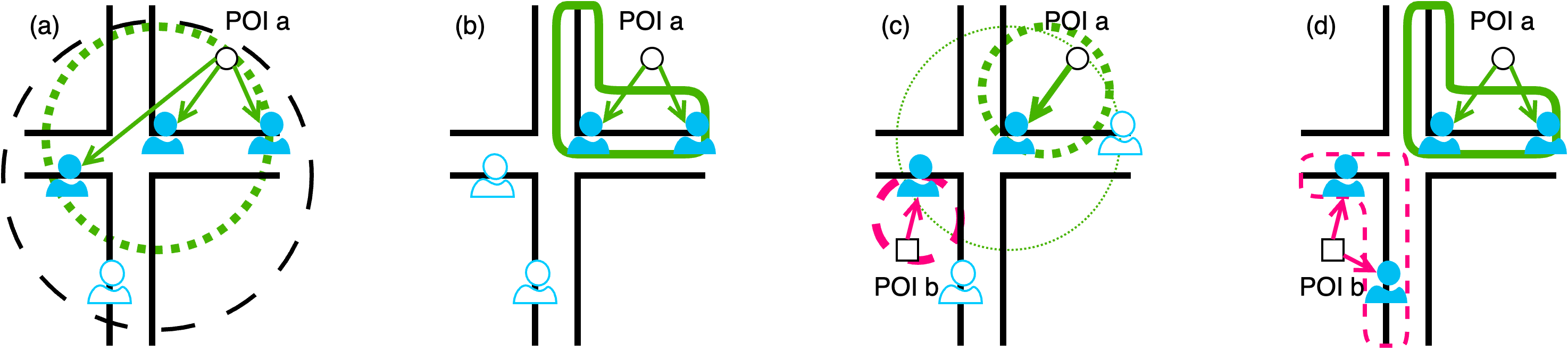}
\caption{Concept comparison of circular geofences and discrete geofences in high-resolution areas (such as urban environments).
Figure~\ref{fig-intro}a and b illustrate information notification using geofences.
Compared with circular geofences in a, discrete shapes in b could form trajectory-oriented regions, which match human mobility data better.
For urban environments with multiple POIs, like in Figure~\ref{fig-intro}c and d, the flexibility of a discrete geofence could fit street-level trajectories better, which might be challenging for circular geofences, as they may overlap.}
\label{fig-intro}
\end{figure}

Traditionally, geofences have been treated as static entities designed by operators or service providers, with predetermined boundaries created in advance based on specific locations or transportation areas. 
Sending relevant information about a POI to users who spend time in these areas, as illustrated in Figure~\ref{fig-intro}a, can facilitate purchases or enable information delivery.
However, human movement patterns around POIs vary significantly, depending on the day and situation.
To effectively utilize human mobility data, designing geofences adaptively, which can be studied as data-driven geofence design problems~\cite{Garg2017, Sasaki2024}, is suitable.
The geofence design problem involves creating adaptive boundaries (fences) that adjust according to the location and context of a user.
Existing geofence design approaches have primarily focused on circular geofences (Circular GeoFences; CGFs).
In Figure~\ref{fig-intro}a, we adjust the circular area to select the users that are relevant to the POI (POI a). 
However, circular modeling lacks the ability to capture flexible shapes or complex spatial boundaries. 
This limitation particularly persists when designing mobility experiences using geofences in urban or high-density environments, where overlapping notification areas may occur or information delivery may fail because of insufficient coverage. 
We then propose using discrete shapes, as illustrated in Figure~\ref{fig-intro}b, to create flexible regions that match human movement data better and are convenient for modeling urban environments with multiple POIs in Figure~\ref{fig-intro}c and Figure~\ref{fig-intro}d. 

This study reframes the geofence design problem within the framework of combinatorial optimization problems involving 0-1 variables to handle complex constraints and represent arbitrary shapes (Figures~\ref{fig-intro}b and d, compared with a and c).
The geofences designed with such flexible spatial boundaries are referred to as \textit{discrete geofences}.
This study aims to overcome the limitations in shape representation inherent in CGFs, as other polygon extraction algorithms or clustering methods, such as DBSCAN~\cite{Ester1996}, have different objectives and assumptions.
Therefore, we evaluate the effectiveness of the discrete geofence design relative to CGFs.

The 0-1 integer programming problems belong to the theoretically NP-hard class of combinatorial optimization problems, exhibiting high computational complexity.
However, recent advances, particularly quantum annealers~\cite{Johnson2011}, have led to the development of solvers~\cite{Yamaoka2015, Tatsumura2019, Matsubara2020} that efficiently and accurately solve combinatorial optimization problems expressed in quadratic unconstrained binary optimization (QUBO) format.
Furthermore, QUBO problem formulation has been actively studied in the field of mathematical optimization, with this approach proving particularly promising for addressing these problems~\cite{Rehfeldt2023}.
So far, this representation has been successfully applied to various practical problems across diverse fields, including polymer phase separation~\cite{Endo2022}, transportation services~\cite{Inoue2021, Otaki2023}, finance~\cite{Ding2023}, and many others~\cite{Yarkoni2022, Hussain2020, Nishimura2019, Ohzeki2019, Syrichas2017, Tabi2021, Weinberg2023, topo2023, Okada2023, Okada2024}.
Therefore, we could achieve both high-precision and high-speed solution performance by effectively formulating optimization problems related to geofences in QUBO or quadratic optimization problems.

The main contributions of this paper are as follows.
First, we formally define the discrete geofence design problem and introduce discrete geofences, which are compared with existing CGFs.
To represent arbitrary shapes, we propose multiple soft constraint terms and develop a method for combining them to implement optimization problems.
Next, we conduct extensive computational experiments using both synthetic and real-world data, comparing performance across several solvers and analyzing the optimized geofences.
Finally, we compare our approach with actual real-world geofence applications and discuss prospects for the proposed method, which can represent complex geometric shapes.

\section*{Methods}
\label{sec-method}

\begin{figure}[h]
\centering
\includegraphics[height=40mm]{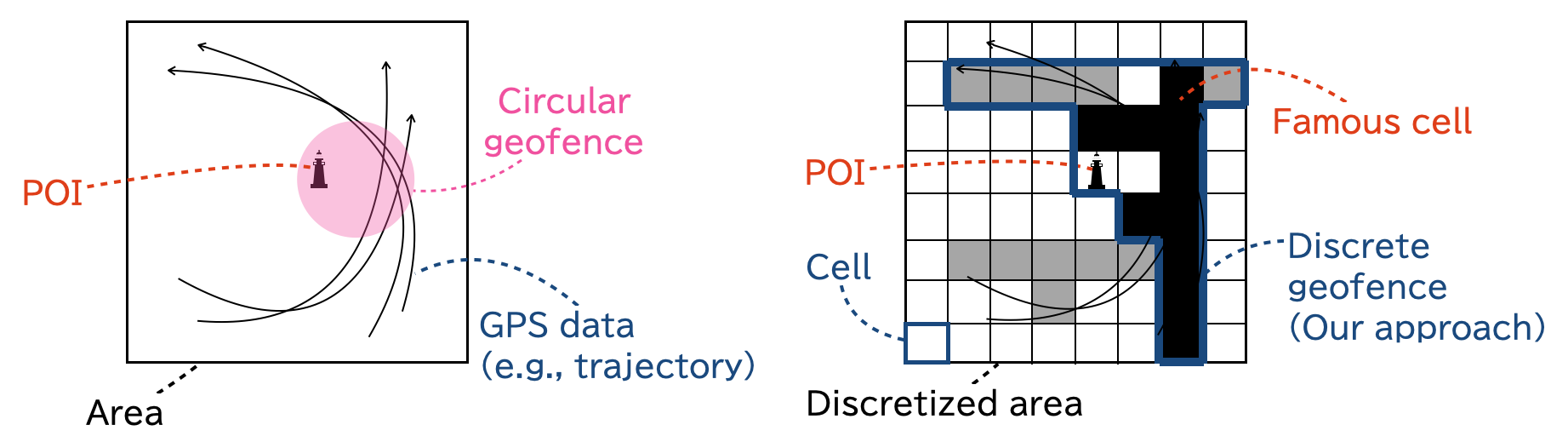}
\caption{Conceptual comparisons: circular and discrete geofences to capture human mobility data (e.g., GPS data) around the target POI.}
\label{fig-concept}
\end{figure}

This section defines the \textit{discrete geofence design problem}, and formulates it as a 0-1 integer programming problem.
Our approach can select appropriate grids after discretizing the area of input GPS data to represent complex shapes.
Figure~\ref{fig-concept} illustrates the differences between the existing circular and our discrete geofences.

\subsection*{Existing Approach: Circular Geofence Design Problems}
\label{subsec-circular-gf}

Following Sasaki et al.~\cite{Sasaki2024}, we consider scenarios where users receive location-based notifications when they enter or exit circular geofences.
The examples include geo-targeted coupon distribution or push notifications for local tourist information.
When setting large geofences for notifications, information may be unintentionally sent to geographically irrelevant users, reducing the effectiveness of the notification.
Conversely, setting extremely small geofences may prevent users from receiving the intended information effectively.
Therefore, developing data-driven methods to establish geofences of appropriate size is necessary for location-based services.

\subsubsection*{Formulation}
\label{subsub-cgf-formulation}

Previous studies have modeled geofences as circular shapes and addressed the geofence design problem as constructing circular shapes around each POI based on the following assumptions and principle.

Let us assume a two-dimensional plane, $\mathbb{R}^2$.
For two points $p, q$ on $\mathbb{R}^2$, we denote the Euclidean and Manhattan distances by $d(p, q)$ and $d^\mathrm{MD}(p, q)$, respectively.
POI $P$ is given to design spatio-temporal events, and its location is represented as $P=(\POIattr{x}, \POIattr{y}) \in\mathbb{R}^2$.
Given are mobility data surrounding a POI; the trajectory data is provided as a set of trajectories, denoted as $\DB=\{T_1, \dots, T_m\}$.
Each trajectory $T_j\in\DB$ consists of a unique user identifier $\uid_j$ and a sequence of points $T_j=\langle\uid_j,\seq_j\rangle$.
Point sequence $\seq_j$ comprises GPS observation points $(t_k,x_k,y_k)$, where each observation records the user's location at $(x_k,y_k)$ at time $t_k$.
A \textit{circular geofence} is represented as a triplet $\GF{i}:=(x_i, y_i, r_i)$, where $(x_i, y_i)$ is the center of the circle and $r_i$ is the radius, as illustrated in Figure~\ref{fig-concept}.
Notably, we denote the distance between point $(x_j, y_j)\in\mathbb{R}^2$ and circular geofence $\GF{i}$ by $d((x_j, y_j), \GF{i}) := d((x_j, y_j), (x_i, y_i))$.
On these data and a parameterized geofence representation, the circular geofence design problem is a multi-objective optimization problem that computes $\GF{i}$ with respect to the evaluation functions for the following two requirements.
\begin{description}
\item[(Requirement 1)] the relevance of the geofence to the target POI, and 
\item[(Requirement 2)] the coverage rate of data $\DB$.
\end{description}
In practice, (Requirement 1) and (Requirement 2) can be formulated as different objective functions according to applications.
For example, Sasaki et al. defined (Requirement 1) as the average of the minimum and maximum distances between designed geofence $\GF{i}$ and the center of the target POI, $P=(\POIattr{x}, \POIattr{y})$, as shown in Eq.~\eqref{eq-cgf-dist}.
\begin{equation}
    \label{eq-cgf-dist}
    f(\GF{i}; P) := \frac{1}{2}\left\{
        d((\POIattr{x}, \POIattr{y}), \GF{i})+r+|d((\POIattr{x}, \POIattr{y}), \GF{i})-r|
    \right\}.
\end{equation}
For the coverage rate, two approaches can be considered: one using unique user IDs for user-level coverage and another using timestamps for point-level coverage.
In the formulation by Sasaki et al., for (Requirement 2), the coverage rate of geofence $\GF{i}$ is defined as the ratio of unique users whose trajectories intersect with the geofence, as shown in Eq.~\eqref{eq-cgf-cover}.
\begin{equation}
    \label{eq-cgf-cover}
    g^\mathrm{cover}(\GF{i}) := \left|
    \{
    \uid_j \mid 
    \exists \langle\uid_j, \seq_j \rangle \in \DB
    \text{ s.t. }
    \exists (t,x,y) \in \seq_j, d((x,y), \GF{i}) < r
    \}
    \right| / |\DB|.
\end{equation}

In this study, we follow the existing method and use the user-level coverage rate.
Using the coverage rate, a minimum coverage rate, $\mathit{cr}_\mathrm{limit}$, with threshold parameter $\mathrm{cr}_\mathrm{limit}$ and coefficient $\mu$ is proposed as Eq.~\eqref{eq-cgf-mincover}.
\begin{equation}
    \label{eq-cgf-mincover}
    g^\mathrm{mincover}(\GF{i}) := \mu\max\left(0, \mathit{cr}_\mathrm{limit} - g^\mathrm{cover}(\GF{i})\right).
\end{equation}
Using these functions, the circular geofence design problem by Sasaki et al.~\cite{Sasaki2024} can be formulated as an optimization problem with the following inputs and outputs:
\begin{quote}
\begin{description}
    \item[Input] A POI $P=(\POIattr{x},\POIattr{y})$ and a trajectory dataset~$\DB$.
    \item[Output] A circular geofence $\GF{i}=(x_i,y_i,r_i)$ that maximizes $f(\GF{i}; P)+g^\mathrm{mincover}(\GF{i})$.
\end{description}
\end{quote}
Please refer to \cite{Sasaki2024} for more details on the definitions and differences in methods.

\subsection*{Baseline Examples with Existing Methods and Identified Challenges}
\label{subsub-cgf-example-and-challenge}

\begin{figure}[htb]
\centering
\includegraphics[height=32mm]{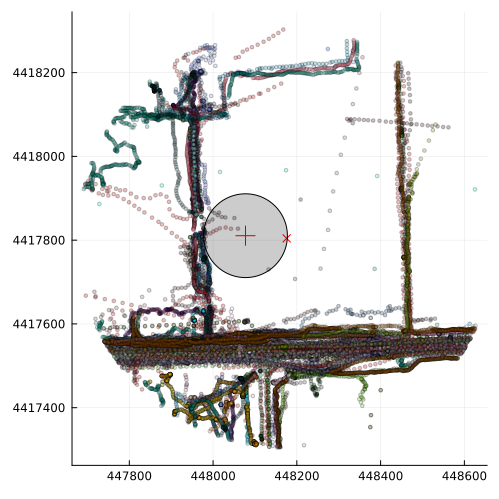}
\includegraphics[height=32mm]{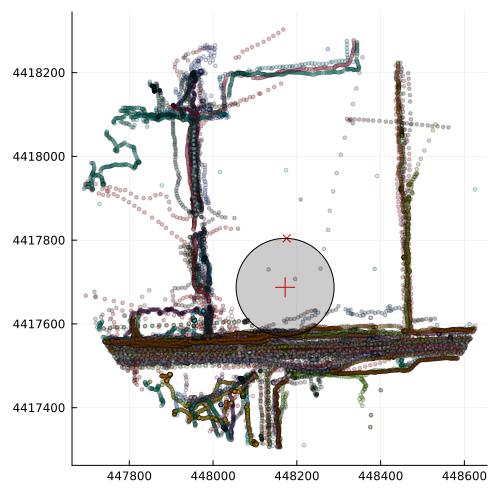}
\includegraphics[height=32mm]{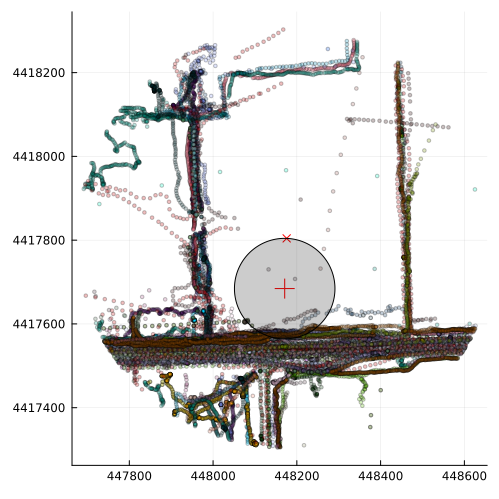}
\includegraphics[height=32mm]{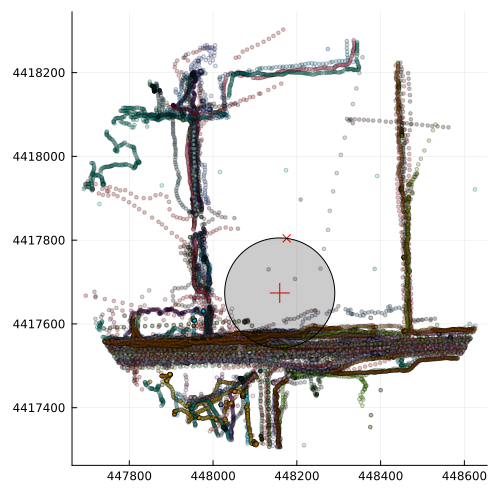}
\includegraphics[height=32mm]{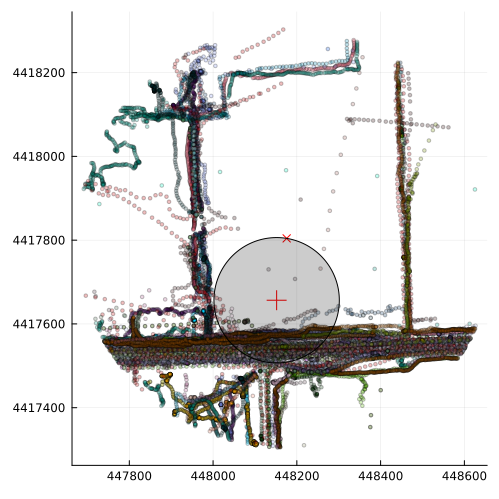}
\caption{
    Examples: parameter $\mathit{cr}_\mathrm{limit}$ is varied $0.1,0.3,0.5,0.7,0.9$ and CGFs are computed. In the figures, the gray circles represent CGFs, and the colors of the data points distinguish unique user IDs.
}
\label{fig-circular-comp}
\end{figure}

To clarify the limitations of the existing circular geofence design approach, this section presents preliminary computational examples using a real-world dataset.
These examples reveal fundamental issues, which we summarize as two main challenges and use as the motivation for our proposed discrete geofence formulation.
For this purpose, we reproduce the computational example of the geofence design problem using the GeoLife dataset\cite{Zheng2008, Zheng2009, Zheng2010}.
Figure~\ref{fig-circular-comp} shows the POIs represented by $\times$, and each GPS-recorded point is indicated by $\circ$, and the computed circular geofences are shown as gray circles.
As shown in Figure~\ref{fig-circular-comp}, the geofences include the specified POIs while encompassing many GPS points, owing to the minimum coverage objective in Eq.~\eqref{eq-cgf-mincover}.

In this paper, we focus on two challenges when reformulating the geofence design problem with discrete shapes.

\paragraph{Challenge 1: Representation Capability}
\label{para-cgf-problem1}

\begin{figure}[htb]
\centering
\includegraphics[height=40mm]{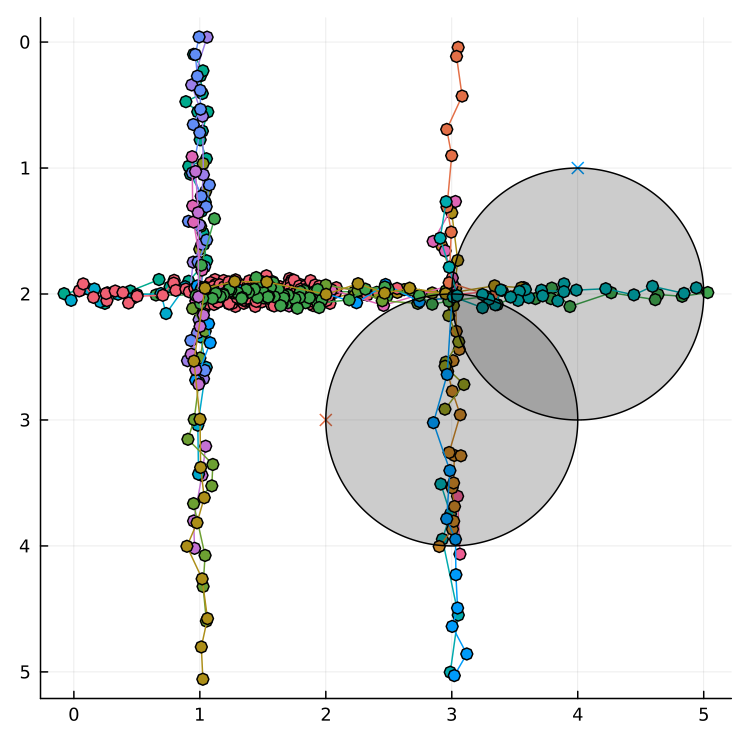}
\includegraphics[height=40mm]{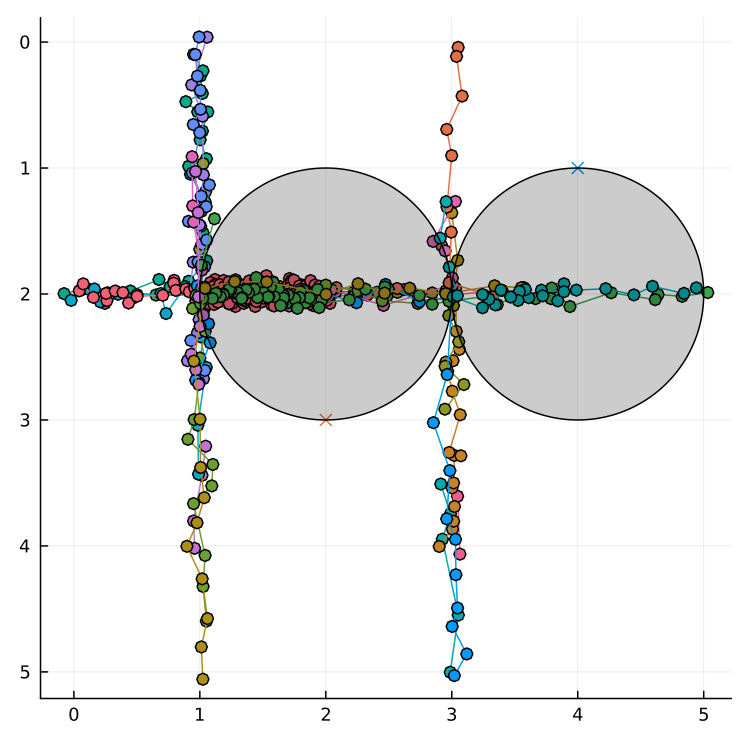}
\caption{Examples. Two circular geofences corresponding to two POIs overlap. To resolve this issue, it is necessary to establish a clear design principle and a modeling approach based on the original formulation.}
\label{fig-modeling-example}
\end{figure}

Evidently, the circular geofences have limitations in representing complex shapes, specifically in urban and high-density environments.
Figure~\ref{fig-modeling-example} shows that circular geofences tend to overlap between POIs (blue and red cross markers), making it difficult to adjust their sizes and cover input data simultaneously.
To overcome this limitation, we propose representing geofences using discrete cells, enabling us to represent arbitrary shapes, as illustrated in Figure~\ref{fig-concept}.

\paragraph{Challenge 2: Objective Function}
\label{para-cgf-problem2}

The current objective function, $f(\GF{i}) + g^\mathrm{mincover}(\GF{i})$, is not a pure multi-objective optimization problem, as it defines the single objective function as a weighted sum of two terms.
As a result, depending on the input data, the optimization may yield solutions that mix different intentions.
Fr example, geofences that extend to distant points unrelated to the target POI.
This occurs because circular shapes, while reasonable in terms of communication energy and distance, inherently favor covering larger areas efficiently, even if they go beyond the intended region.

To address this limitation, we introduce a discrete modeling approach in which geofences are represented at the cell level using 0-1 integer programming.
This formulation allows flexible designs, as any cell can be included in the solution.
However, such flexibility also brings new challenges: (i) the optimized shapes may become fragmented, and (ii) area costs and coverage values may become intertwined with distance dependencies. To overcome these issues, we propose weighting both terms by the Manhattan distance from the POI, thereby decomposing the objective function into two clearly interpretable components: communication cost and information value.

\subsection*{Our Approach: Discrete Geofence Design Problems}
\label{subsec-dgf}

When using high-precision location data or constructing geofences in dense urban environments, circular geofences, even when optimized, may produce coarse representations that result in irrelevant or low-relevance notifications.
We address this limitation by introducing the \textit{discrete geofence design problem}, which enables representation of complex geometric shapes.
By utilizing geofences with complex shapes, we could deliver targeted information only to users performing specific actions and add more nuanced motivations for movement.

\begin{figure}[htb]
\centering
\subcaptionbox{Normalized input\\(red:POI, black:GPS)\label{fig:l1}}[0.22\columnwidth]{
    \includegraphics[width=\linewidth]{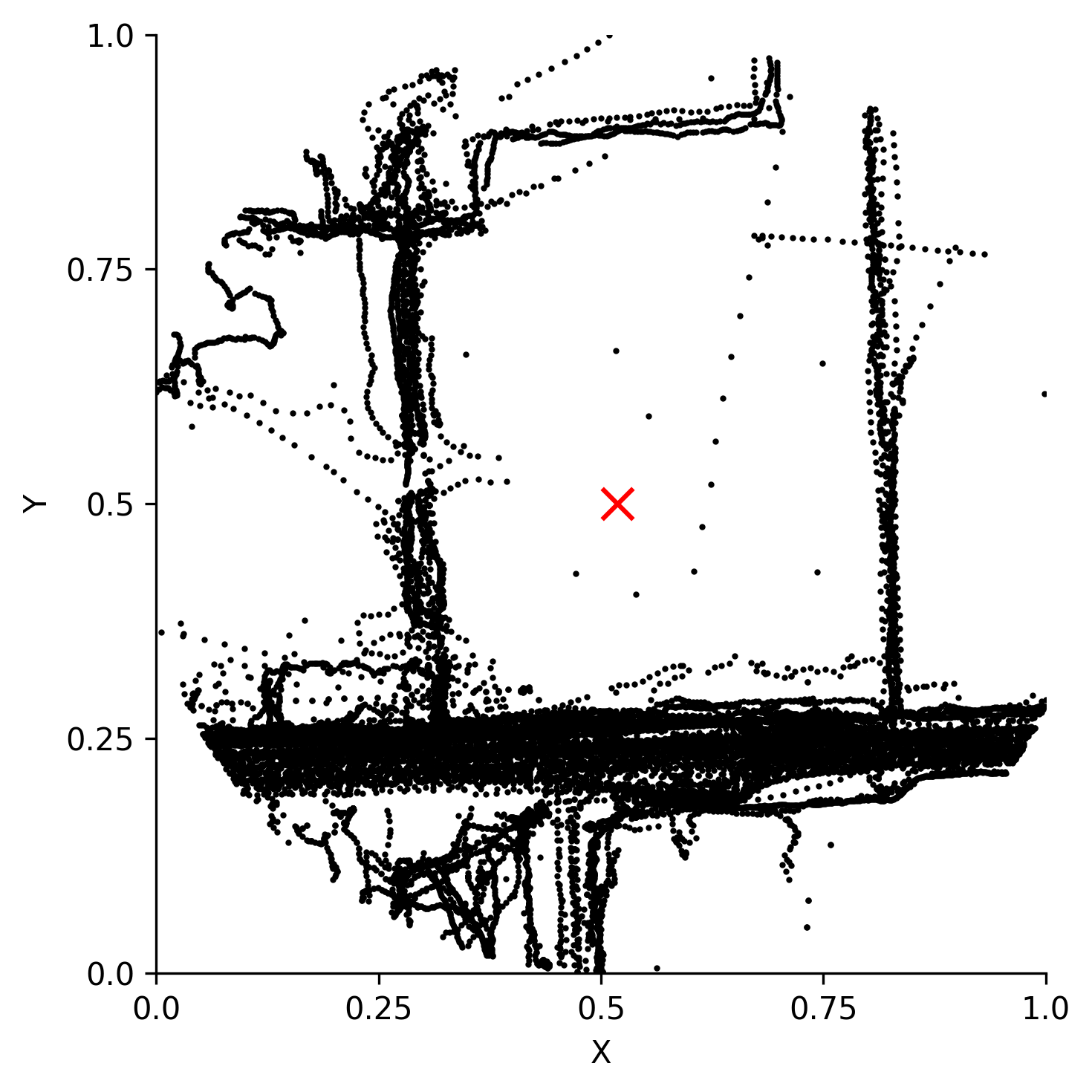}
}
\subcaptionbox{Discretized input ($d=3$)\label{fig:l2}}[0.24\columnwidth]{
    \includegraphics[width=\linewidth]{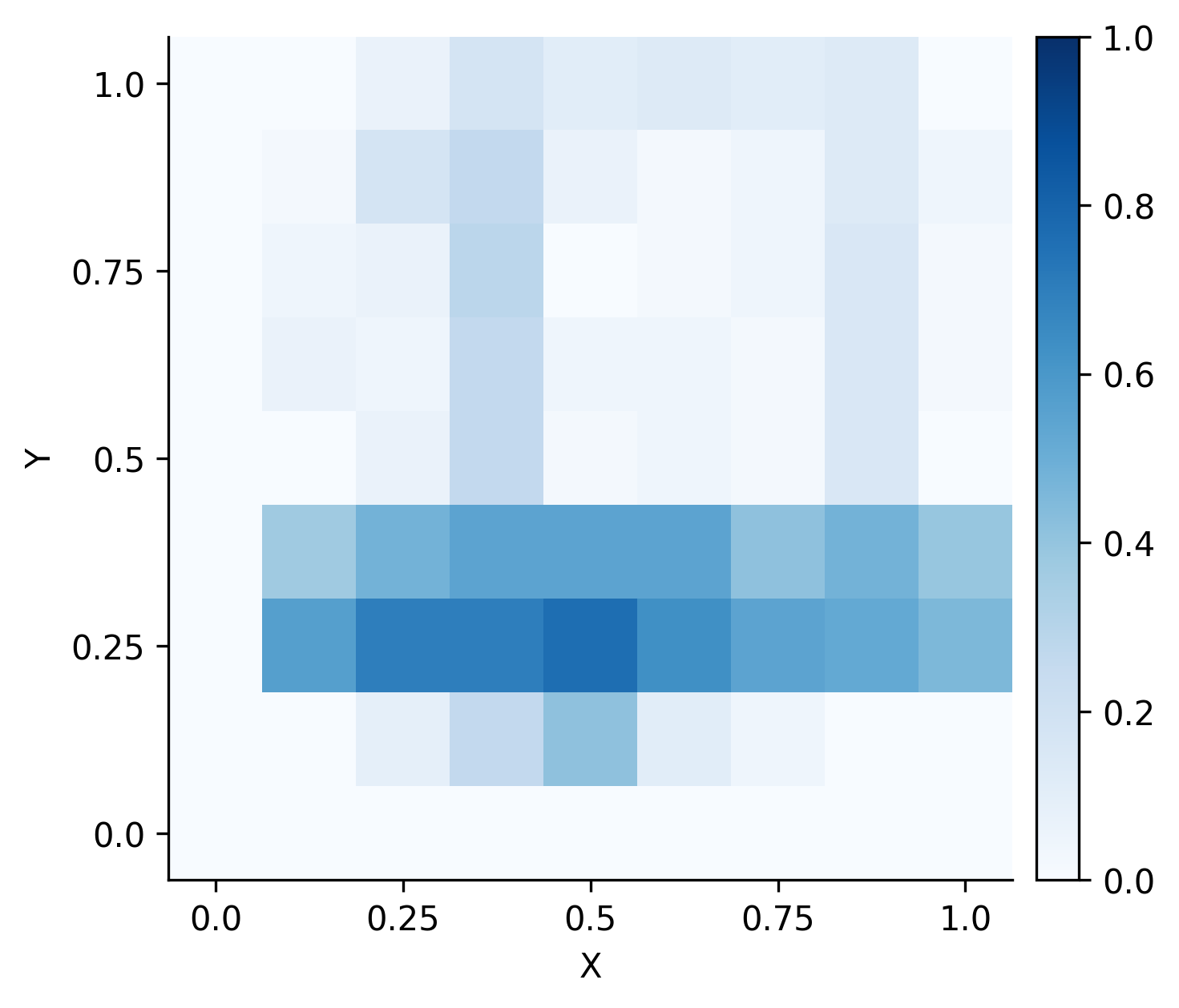}
}
\subcaptionbox{Case ($d=4$)\label{fig:l3}}[0.24\columnwidth]{
    \includegraphics[width=\linewidth]{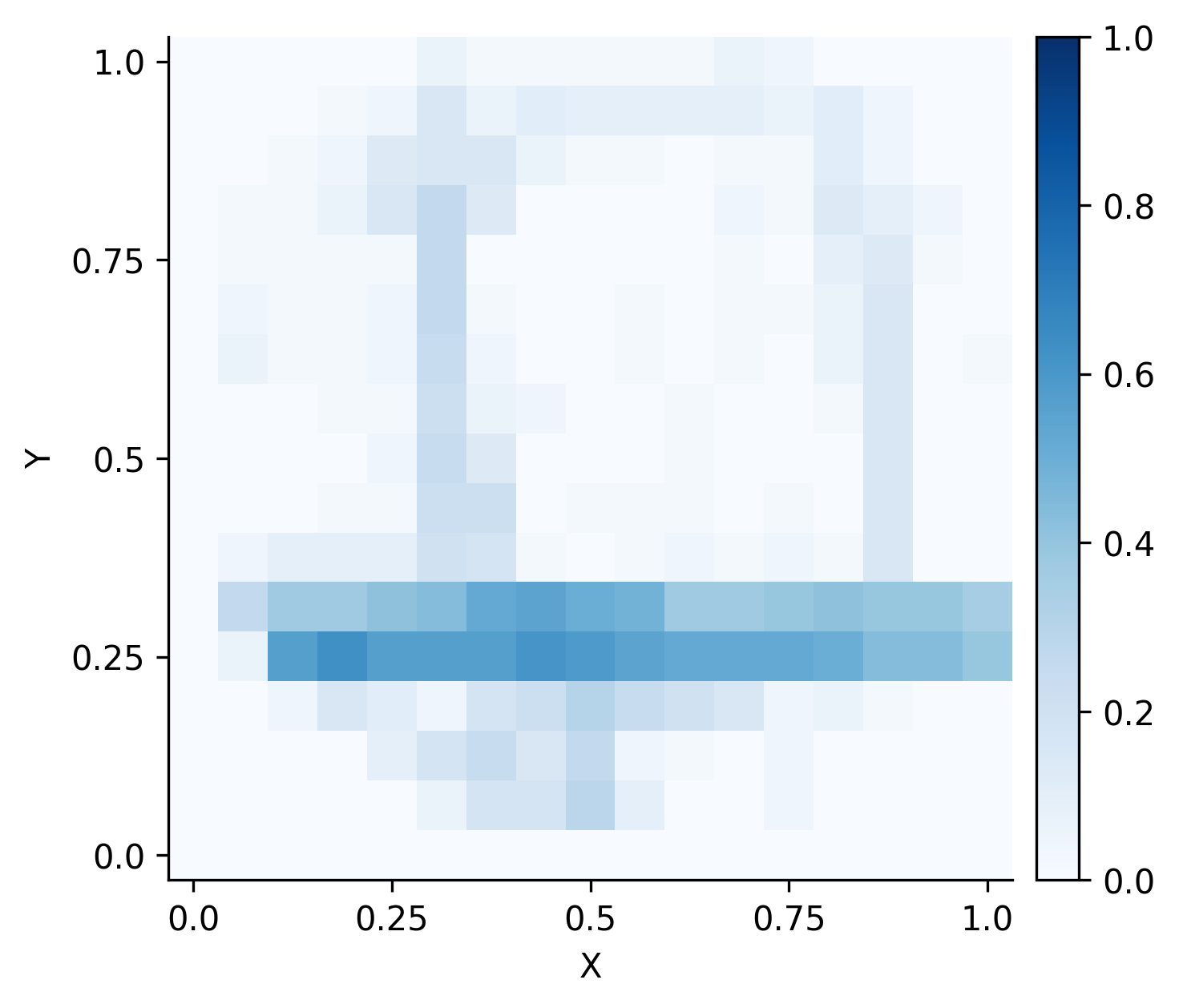}
}
\subcaptionbox{Case ($d=5$)\label{fig:l4}}[0.24\columnwidth]{
    \includegraphics[width=\linewidth]{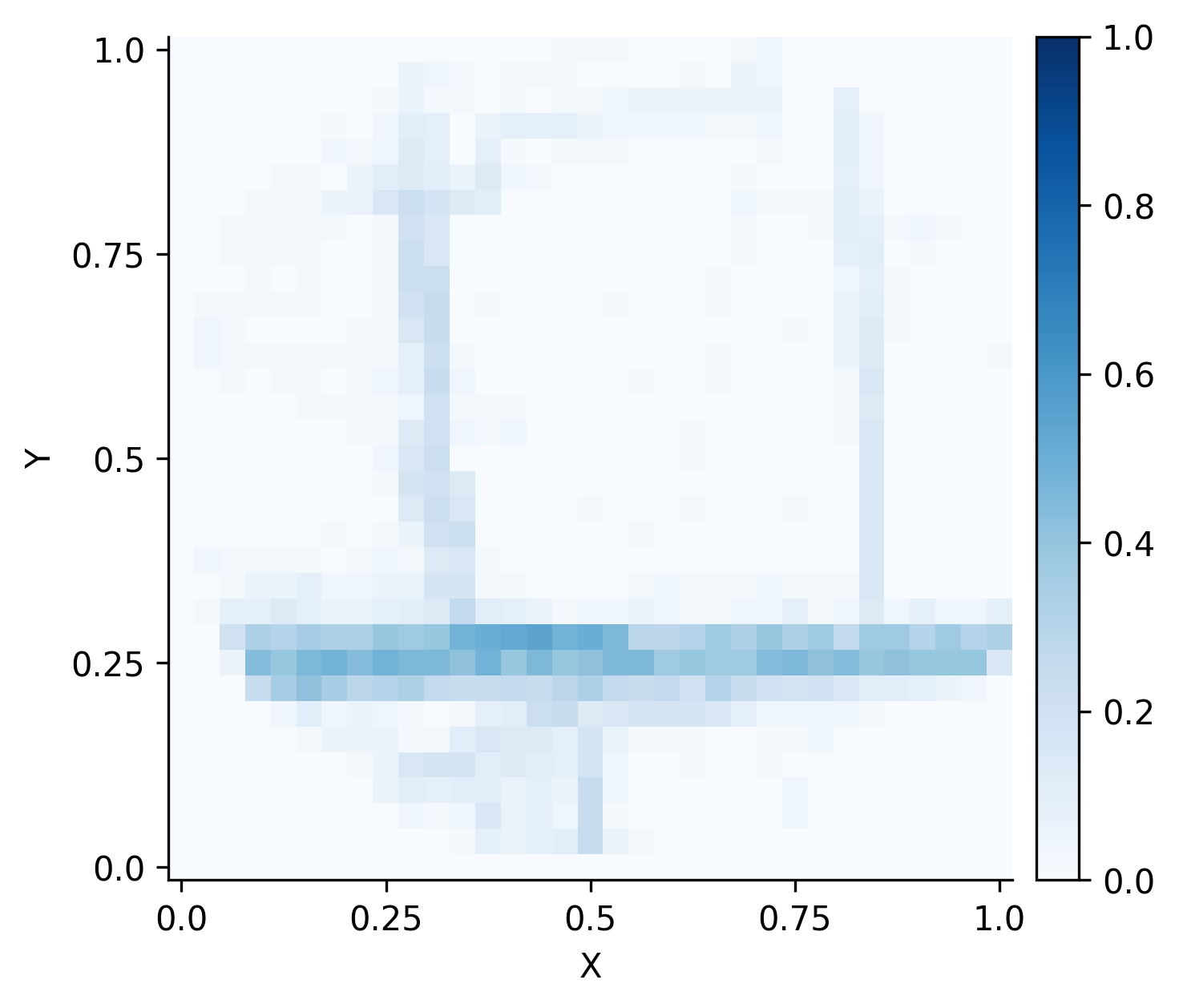}
}
\caption{Example of data preprocessing: Depending on the number of bits used for computation, we can represent shapes with fine details (deep levels) to coarse shapes (low levels). We demonstrate examples for $d=3,4,5$. By discretizing and counting the number of grid-based user IDs, we obtain the discretized user data matrix, $V$; the shading in each cell $(i, j)$ corresponds to $V_{i,j}\in[0,1]$.
}
\label{fig-input_data}
\end{figure}

Throughout the following discussion, we assume the input data has been normalized, transforming both the $x$- and $y$-axes value ranges to the interval $[0, 1]$.
As illustrated in Figure~\ref{fig-concept}, we address the problem of selecting appropriate cells after discretizing the area of input GPS data to represent complex shapes.
For notation simplicity, we assume the discretization level parameter, $d \geq 1, d \in \mathbb{N}_+$, and discretize both axes into $2^d =: L$ levels.
Figure~\ref{fig-input_data} illustrates an example of the Geolife dataset used in Figure~\ref{fig-circular-comp}, where the data is discretized into cells for $d=3, 4, 5$.
Once $d$ increases, the input data can be represented in finer shapes, and it is evident that the problem of designing discrete geofences can be flexibly accommodated by setting $d$.

Under the above assumptions, we propose the discrete geofence design problem as an optimization problem with the following inputs and outputs.
\begin{quote}
\begin{description}
    \item[Input] A POI $(P_r, P_c)$, where $P_r, P_c \in [L]:=\{1\,\dots,L\}$, and a two-dimensional data matrix $V\in[0,1]^{L\times L}$.
    \item[Output] A binary two-dimensional matrix $X\in\mathcal{X}:=\{0,1\}^{L\times L}$ representing a discrete geofence (where $X_{i,j}=1$ indicates that grid position $i, j\in [L]$ is contained within the geofence, and $\mathcal{X}$ represents the set of all possible decision variable matrices $X$).
\end{description}
\end{quote}

\noindent
For the decision variable matrix, $X\in\{0,1\}^{L\times L}$, we consider regions where $X_{i,j}=1$ as geofences.
In this study, we directly generalize the properties of circular geofences and consider extracting the connected regions that should receive notification about the input POI location $(P_r, P_c)$.

\subsection*{Formulations: From 0-1 Integer Programming to QUBO and Constrained Variants}
\label{sub-dgf-qubo}

This section aims to find discrete geofences using QUBO solvers by formulating the optimization problem from the previous section in a QUBO formulation. 

\subsubsection*{Core Objective Function: Area and Coverage}
\label{subsub-dgf-objective}

We define two objective functions to satisfy (Requirement 1) and (Requirement 2) that arise from circular modeling.

\begin{equation}
    \label{eq-dgf-area}
    f_\mathrm{area}(X) := \sum_{i, j} X_{i, j}.
\end{equation}

\begin{equation}
    \label{eq-dgf-cover}
    f_\mathrm{cover}(X) := \sum_{i, j} C_{i, j} V_{i, j} X_{i, j}.
\end{equation}
In Eq.~\eqref{eq-dgf-area}, $f_\mathrm{area}(X)$ represents the number of selected cells, extending Eq.~\eqref{eq-cgf-dist}.
When extending the circular geofence design problem, to ensure that the discrete geofence is designed adaptively, $f_\mathrm{cover}(X)$ should be sufficiently large for information notifications, while $f_\mathrm{area}(X)$ representing the geofence area should be minimized completely to reduce irrelevant notifications.

Ensuring that $f_\mathrm{cover}(X)$ remains sufficiently large to inform many users around POIs is important in discrete geofences.
Consequently, Eq.~\eqref{eq-dgf-cover} with $C_{i,j}=1$ corresponds to a binary-variable representation of Eq.~\eqref{eq-cgf-cover}, which can be adopted in our modeling.
Notably, it is more challenging to define Eq.~\eqref{eq-cgf-mincover} used in the original formulation as the function of the 0-1 variables; thus, we focus on the covering objective in Eq.~\eqref{eq-dgf-cover}.
Our modeling is flexible to consider POIs; for example, if we set $C_{i,j}$ as a value inversely proportional to the Manhattan distance from target POI $P=(P_r, P_c)$, we can model the relevance between input data $V_{i,j}$ with POI $P=(P_r, P_c)$ in the covering objective.
That is,
\begin{equation}
    \label{eq-weighted-cover}
    C_{i,j} = \left(1 + d^\mathrm{MD}((P_r, P_c), (i, j))\right)^{-\alpha}.
\end{equation}
With parameter $\alpha > 0$, we can create an objective function that represents the extent of influence considering the location of the POI.
This setup allows us to incorporate the intention that users at greater distances from the POI have lower coverage value; that is, covering locations closer to the POI enables more relevant information notifications.
In the sections below, we compute discrete geofences by adjusting the weight coefficients applied to both $f_\mathrm{cover}(X)$ and $f_\mathrm{area}(X)$.

\subsubsection*{Core Constraints: Connectivity of Discrete Regions and the input POI}
\label{subsub-dgf-constraint}

This section explains the components of our QUBO formulation that shape discrete regions.
We introduce two constraints based on the Domain Wall constraint and the adjacency restriction between cells, which represent an arbitrary geofence using cell-wise discrete decision variables.

\paragraph{Two-dimensional Domain Wall Constraints}
\label{para-dgf-2dw}

\begin{figure}[htb]
\centering

\subcaptionbox{Boolean conjunction to represent an island in one-dimensional case.\label{fig-1dw}}[\linewidth]{
    \includegraphics[height=40mm]{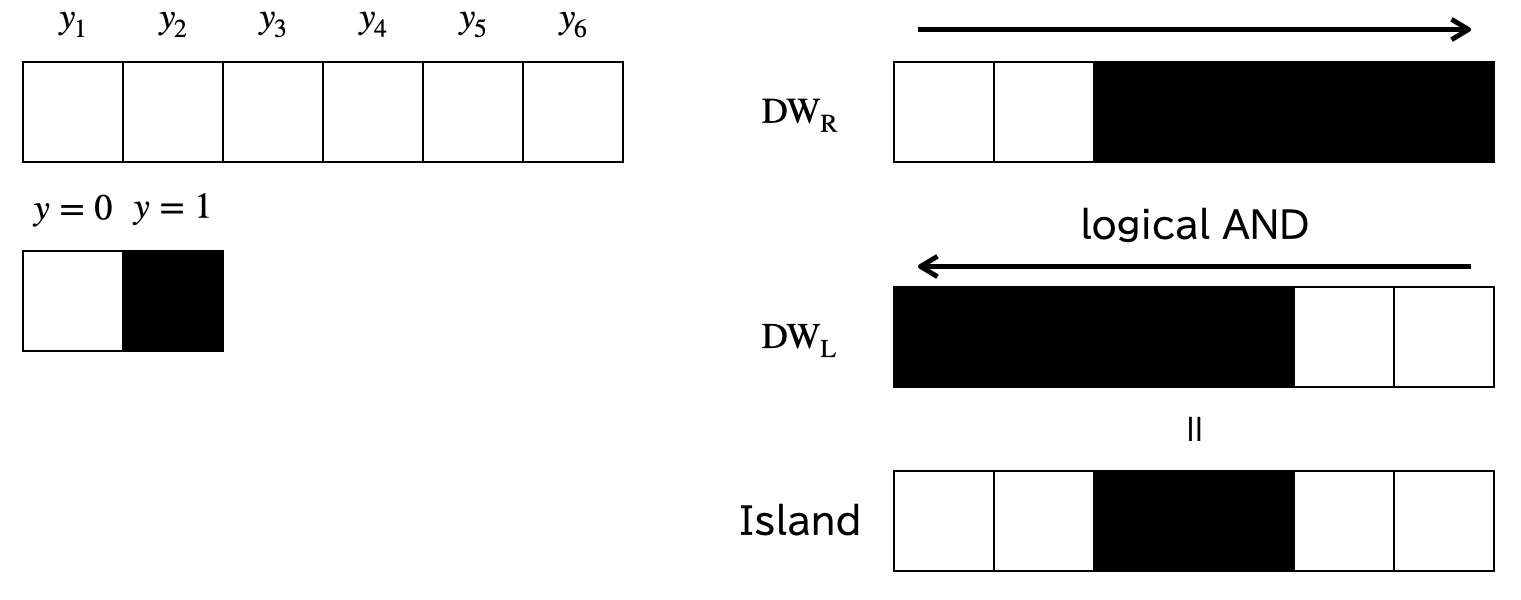}
}

\vspace{1em}

\subcaptionbox{Two-dimensional example.\label{fig-2dw}}[\linewidth]{
    \includegraphics[height=40mm]{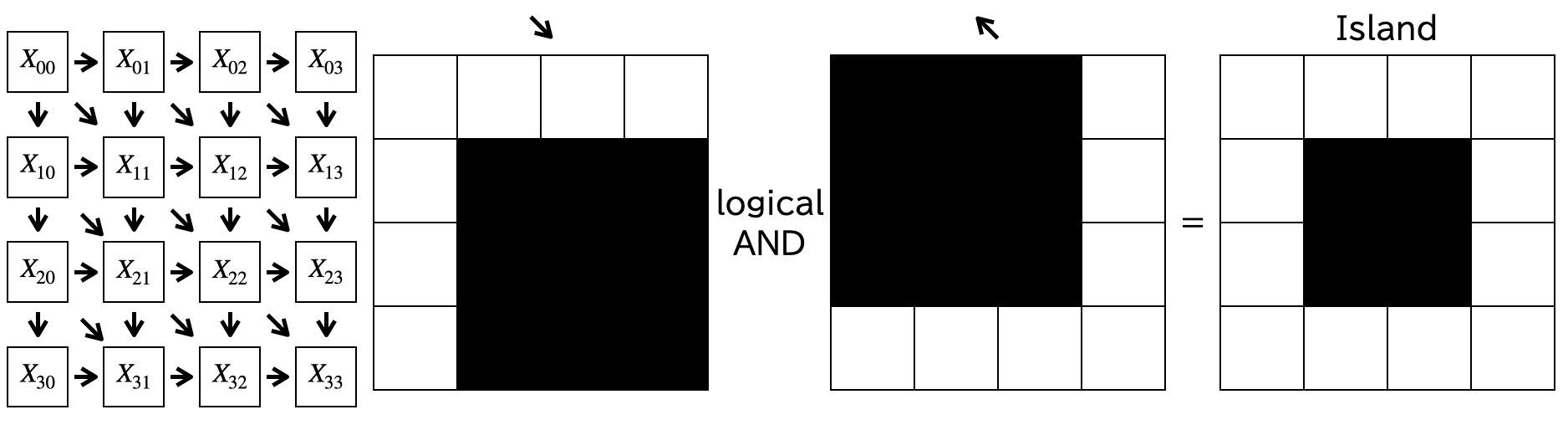}
}

\caption{Domain Wall constraints to model discrete geofences.}
\label{fig-dw-islands}
\end{figure}

For a bit string of length $N$ denoted by $y\in\{0,1\}^N$, a representation known as the Domain Wall constraint~\cite{Chancellor2019}, which requires that the bit string, $y$, must consist of zero or more consecutive zeros followed by zero or more consecutive ones.
The constraint for one-dimensional cases is expressed by $\sum_{i=0}^{N-2} y_i(1-y_{i+1}) = 0$.
Bit strings satisfying the Domain Wall constraint never change from 1 to 0 within the sequence, effectively representing a one-dimensional wall that adheres to the right side of the sequence.
By applying the constraint in both left-to-right ($\mathrm{DW}_\mathrm{R}$) and right-to-left directions ($\mathrm{DW}_\mathrm{L}$), and taking the logical AND ($\wedge$ in bit operation) of the two walls, as illustrated in Figure~\ref{fig-1dw}.

This paper extends the one-dimensional Domain Wall constraint to represent discrete island structures.
Specifically, we extend the one-dimensional formulation in both the right-downward ($\mathrm{RD}$) and left-upward directions ($\mathrm{LU}$), as illustrated in Figure~\ref{fig-2dw}.
We explain the implementation for the right-downward direction below.
For a given cell, $(r,c)$, we examine the bits in its right cell $(r,c+1)$, bottom-right cell $(r+1,c+1)$, and bottom cell $(r+1,c)$, and impose constraints analogous to those in one-dimensional Domain Wall constraints.
We define the following Eq.~\eqref{eq-2dw}:
\begin{equation}
    \label{eq-2dw}
    f_\mathrm{2DW}^\mathrm{RD}(X) := \sum_{r=0}^{L} \sum_{c=0}^{L} \sum_{(r',c')\in\mathcal{N}_\mathrm{RD}(r,c)} X_{r,c}(1-X_{r',c'}).
\end{equation}

\noindent
Here, $\mathcal{N}_\mathrm{RD}(r,c)$ is a function that returns the indices of cells to the right, bottom-right, and below without exceeding the boundary. If $X$ satisfies all constraints, then $f_\mathrm{2DW}^\mathrm{RD}(X)=0$.
Conversely, when handling the left-upward direction, we use auxiliary function $\mathcal{N}_\mathrm{LU}(r,c)$ that returns the indices of cells to the left, left-up, and above, and similarly define the penalty term $f_\mathrm{2DW}^\mathrm{LU}$.
Through analogous reasoning, we can also consider the reverse diagonal directions (extension to the right-upward RU and left-downward LD directions).
In this paper, we denote the two-dimensional Domain Wall constraint term as $f_\mathrm{2DW}(X)$.

\paragraph{Adjacent Cell Constraint}
\label{para-dgf-ng}

While the domain wall constraints promote global connectivity of the geofence, they may still allow locally fragmented or irregular boundaries.
To complement this, we introduce adjacency-based constraints, which encourage local smoothness and prevent discontinuities at the cell level.

As an approach for modeling geofence connectivity using $X$, we consider leveraging quadratic terms in the decision variables and incorporating soft constraints with penalty terms, assuming the use of a quadratic optimization solver.
For two adjacent cells, $(r, c), (r', c')\in [L]\times [L]$, we propose a soft constraint that prevents the formation of geofence cross-sections when incoming values $V_{r,c}$ and $V_{r',c'}$ are close together, thereby maximizing island connectivity.
Using parameter $\sigma>0$, we define coefficient $Q_{(r,c),(r',c')}$ according to Eq.~\eqref{eq-coeff}.
\begin{equation}
\label{eq-coeff}
Q_{(r,c),(r',c')} := \exp \left(-\frac{||V_{r,c} - V_{r',c'}||^2}{2\sigma^2}\right),
\end{equation}
and we define new constraint term $f_\mathrm{ng}(X)$ according to Eq.~\eqref{eq-ng}.
\begin{equation}
\label{eq-ng}
f_\mathrm{ng}(X) := \sum_{r=0}^{L} \sum_{c=0}^{L} \sum_{(r',c')\in\mathcal{N}(r,c)} Q_{(r,c),(r',c')} X_{r,c} (1 - X_{r',c'}).
\end{equation}

\noindent
Here, $\mathcal{N}(r,c)$ is an auxiliary function that returns the indices of valid neighboring cells for cell $(r, c)$.
When considering a 4-neighborhood, this includes cells above, below, to the left, and to the right.
When considering an 8-neighborhood, it includes diagonally adjacent cells along with the 4-neighbors.
Notably, in the following discussions, neighborhood $\mathcal{N}(\cdot)$ was implemented as the 4-neighbors format in Eq.~\eqref{eq-ng}.
This constraint puts penalties when two adjacent cells have different 0-1 values; therefore, the computed geofence can be smoothly connected.

\paragraph*{Relation between POIs and Discrete Geofences}
\label{para-dgf-poi}

Circular geofences often have an implicit physical constraint (e.g., an antenna for notifications and notifications using boundaries), which imposes a hard constraint on the POI.
In contrast, discrete geofences could be managed by modern location-based technologies (e.g., \cite{Bluedot2025}) and can represent any-shaped regions that are not strongly tied to POIs.
According to these different assumptions, we need to prepare different constraints on POIs.
The requirement is to include the POI within the region, as illustrated in Figure~\ref{fig-circular-comp}; we can employ the hard constraint $X_{P_r, P_c}=1$.
Conversely, coefficients $C_{i,j}$ used in objective function $f_\mathrm{cover}(X)$ depend on $(P_r, P_c)$ as Eq.~\eqref{eq-weighted-cover} with power-weighting coefficients $\alpha > 0$, cells distant from the POI do not contribute to coverage; thus, there is no need to include constraints that consider the distance to the POI.
That is, this formulation implicitly incorporates the relationship with the POI, and this constraint can be optional when utilizing the system for cell-based information dissemination, unlike circular geofences.
Assuming the use of solvers capable of handling quadratic expressions, we propose a method for creating regions that satisfy connectivity as closely as possible using soft constraints.

\subsubsection*{Overall Optimization Formulations}
\label{subsub-dgf-optimization}

First, we define the entire optimization formulation in a QUBO formulation.
The discrete geofences we seek should encompass POIs while maintaining maximally connected regions.
When computing soft constraints using the auxiliary functions, we can adjust coefficients $\{A_\mathrm{area}, A_\mathrm{cover}, A_\mathrm{2DW}, A_\mathrm{ng}\}$ to design the desired region, $X$, by minimizing the objective function Eq.~\eqref{eq-obj-total}.

\begin{equation}
\label{eq-obj-total}
\min_{X\in\mathcal{X}} \:\:
A_\mathrm{area} f_\mathrm{area}(X)
- A_\mathrm{cover} f_\mathrm{cover}(X) 
+ A_\mathrm{2DW} f_\mathrm{2DW}(X)
+ A_\mathrm{ng} f_\mathrm{ng}(X).
\end{equation}

\noindent
For pure QUBO-oriented solvers, such as simulated annealing or quantum annealing, which cannot handle hard constraints, we adjust the coefficients and solve the problem using Eq.~\eqref{eq-obj-total} to obtain the solution.
However, each term has a different value range, making the entire computing process of discrete geofence design problems challenging because misaligned coefficients easily generate trivial solutions, like selecting all cells.

To mitigate the difficulty of scaling coefficients, we propose a modified version that aims to control the size of $X$ as hard constraints when using a mathematical optimization solver.
For example, as in Eq.~\eqref{eq-obj-total-hard}, the problem can be formulated with the following hard constraint on the range of cell selection values for discrete geofences $[A^\mathrm{min}_\mathrm{area}, A^\mathrm{max}_\mathrm{area}]$, and solved using a mathematical optimization solver to determine the discrete geofence $X$ with adjusting $\{A_\mathrm{cover}, A_\mathrm{2DW}, A_\mathrm{ng}\}$:

\begin{align}
\label{eq-obj-total-hard}
\min_{X\in\mathcal{X}} &\:\:
- A_\mathrm{cover} f_\mathrm{cover}(X)
+ A_\mathrm{2DW} f_\mathrm{2DW}(X)
+ A_\mathrm{ng} f_\mathrm{ng}(X), \nonumber \\
&\text{s.t.\:\:\:}  {A}^\mathrm{min}_\mathrm{area} \leq f_\mathrm{area}(X)  \leq {A}^\mathrm{max}_\mathrm{area}.
\end{align}

\noindent
In our experiments, we adopt the above area-constrained optimization formulation to control the area of computed discrete geofences, which is comparable to the baseline circular geofences.

\section*{Experiments and Results}
\label{sec-results}

We apply the proposed method to geofence design problems on synthetic and real data.
By comparing the optimized results with those obtained using optimally designed circular geofences, we demonstrate the properties and characteristics of the proposed approach.

\subsection*{Data and Solver}
\label{subsec-environment}

\paragraph{Synthetic Data}
\label{para-setup-synthetic-data}

We utilize artificial trajectory data generated on a planar grid to simulate urban environments, as illustrated in Figure~\ref{fig-modeling-example}.
The simulated data were generated through the following procedure:

\begin{enumerate}
    \item Create a grid graph of size $N$ and thin out its edges to mimic the graph structure of a highly dense urban environment.
    \item Randomly place POIs on a discretized two-dimensional plane.
    \item Define starting point $O$ and destination $D$ on the grid graph, then sample and compute a path from $O$ to $D$ to obtain a sequence of grid vertices. From the vertex sequence, we generate movement series $(x,y)$ with added Gaussian noise to simulate GPS measurement errors.
    \item Build a synthetic dataset $\DB$ by collecting all simulated movements.
\end{enumerate}

\paragraph{Real-World Data}
\label{para-setup-real-data}

We utilized the GeoLife dataset \cite{Zheng2008, Zheng2009, Zheng2010}, which was also employed in circular geofence experiments~\cite{Sasaki2024}.
For preprocessing, we extracted data from GeoLife within the geographic range of latitude $[32.0, 48.0]$ and longitude $[114.0,120.0]$, projecting it onto the Universal Transverse Mercator zone 50. We then retained only the area centered on the worker's cultural palace's (WCP) coordinates $(x,y)=(448175.7, 4417804.3)$ in zone 50, restricting the search radius to $500$ [m].
Subsequently, from the resulting spatially constrained trajectory data, we retained only users with at least 100 GPS points.
Figure \ref{fig:l1} and Table \ref{table-data-all} present an overview of the dataset used.

\begin{table}[htb]
\caption{Complete dataset containing GPS points within 500 [m] of workers' cultural palace (WCP) from the GeoLife dataset}
\label{table-data-all}
\centering
\begin{tabular}{ll}
\toprule
Parameter & Value  \\
\midrule
Number of GPS points & 22,010 \\ 
Number of unique user IDs & 46 \\
Average/standard deviation of GPS points per user & 478.4/629.4 \\
Range of GPS points per user & $[1, 2556]$ \\
Range of $X$ values & $[447691, 448627]$ \\
Range of $Y$ values & $[4417304, 4418303]$ \\
\bottomrule
\end{tabular}
\end{table}

\paragraph{Solvers}
\label{para-setup-solvers}

To characterize the properties and characteristics of the discrete geofence design problem, we solved an optimization problem with different hyperparameters, and the area constraint was set as a hard constraint.
For circular geofences, optimization was performed using the optimization package Metaheuristics.jl in the Julia programming language, similar to the environment used in Figure~\ref{fig-circular-comp}.
For discrete geofences, we employed SCIP~\cite{BolusaniEtal2024OO} as our solver owing to its capability to handle quadratic terms, as shown in Figure~\ref{fig-results1-input-overlay}.
The SCIP solver was also accessed using the JuMP package in Julia.
Standard parameter settings were similar to those in Figures~\ref{fig-circular-comp} and \ref{fig-results1-input-overlay}, respectively, and varied parameters were explicitly given if they existed.

\subsection*{Experimental Results}
In the following sections, we numerically examine the discrete geofence design problem by systematically varying parameters and problem configurations.
For coverage term $f_\mathrm{cover}$, we employed a weighted function calculated using the Manhattan distance from POI $(P_r, P_c)$:
\begin{equation}
\label{eq-dgf-weighted-cover}
f_\mathrm{cover}(X) := \sum_{i, j} \left(1 + d^\mathrm{MD}((i, j), (P_r, P_c))\right)^{-\frac{1}{2}} V_{i, j} X_{i, j}.
\end{equation}

\subsubsection*{Real-Data Case}
\label{subsub-results-dgf-example}

\begin{figure}[htb]
\centering
\includegraphics[height=40mm]{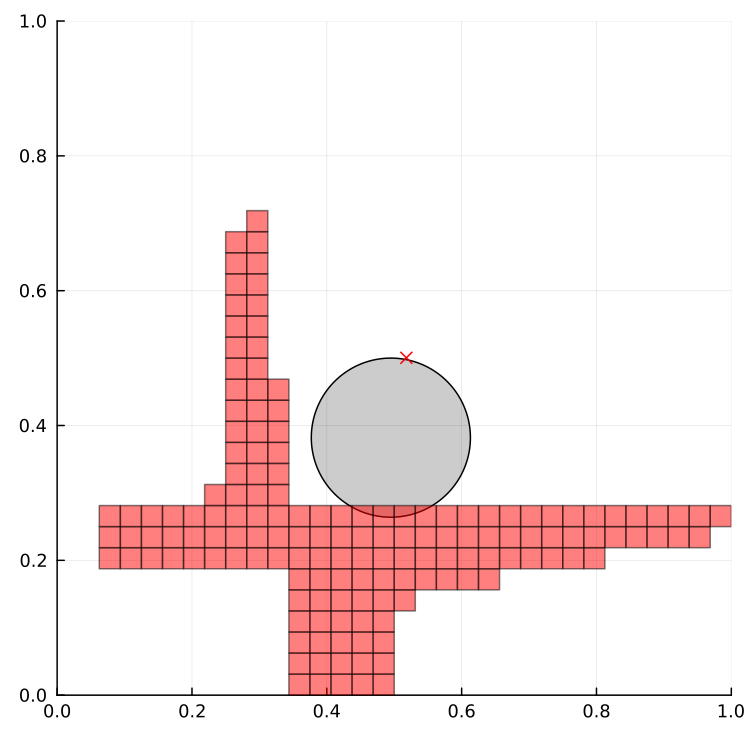}
\caption{Computed circular and discrete geofences with a POI ($\times$).}
\label{fig-results1-input-overlay}
\end{figure}

We first present computational results for real-world data.
Figure~\ref{fig-results1-input-overlay} shows the optimized circular geofence (the center circle, filled in gray) and the optimized discrete geofence (rectangles filled in red with cell edges), where $A_\mathrm{cover}=60.0, A_\mathrm{2DW}=1.0, A_\mathrm{ng}=1.0$ are specified while limiting the number of cells to less than 15\% of the total with discretization level $d=5$.
These results contrast the properties of circular and discrete geofences; thus, we investigate them in the following experiments.

\paragraph{Parameter Effects}
\label{para-exp1-parameters}

\begin{figure}[htb]
\centering
\includegraphics[width=0.8\linewidth]{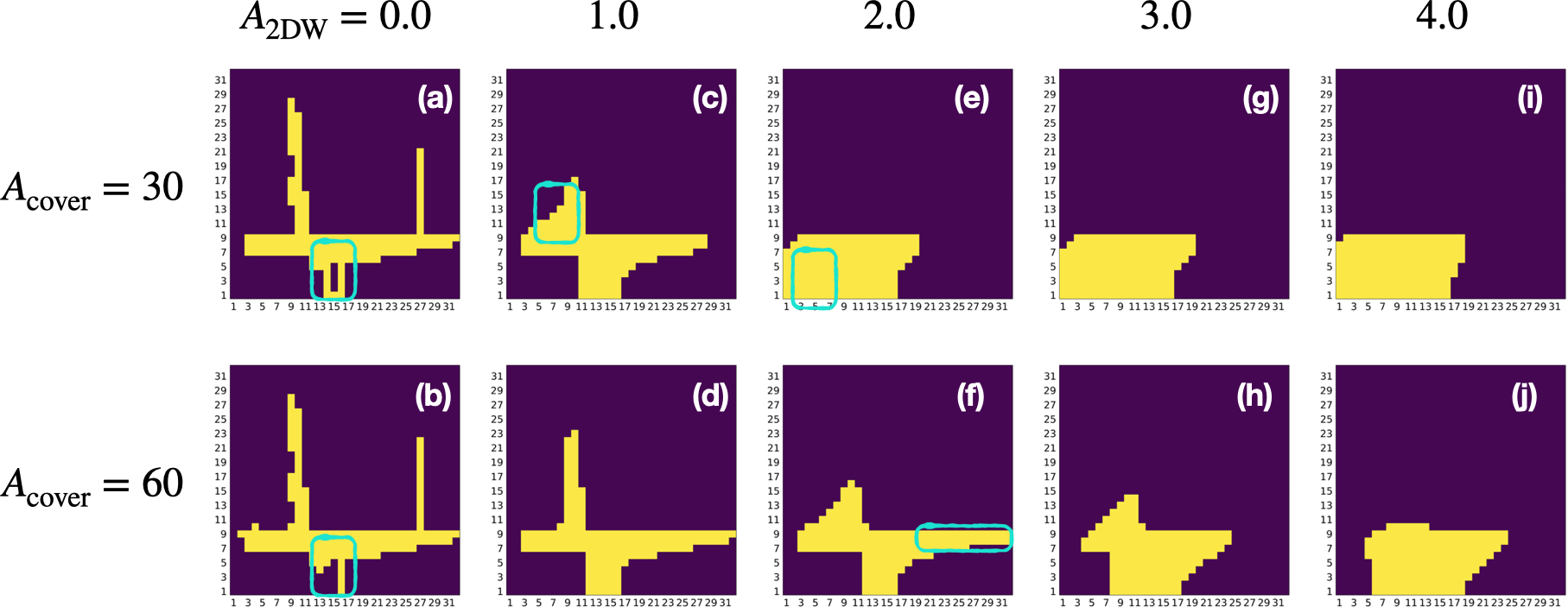}
\caption{Example computation: How the shape of discrete geofences changes when varying $A_\mathrm{2DW}$ with different primary weight $A_\mathrm{cover}\in\{30,60\}$. The effect of adjusting the details of boundaries, which are highlighted by solid lines.}
\label{fig-results2-A2D}
\end{figure}

\begin{figure}[htb]
\centering
\includegraphics[width=0.8\linewidth]{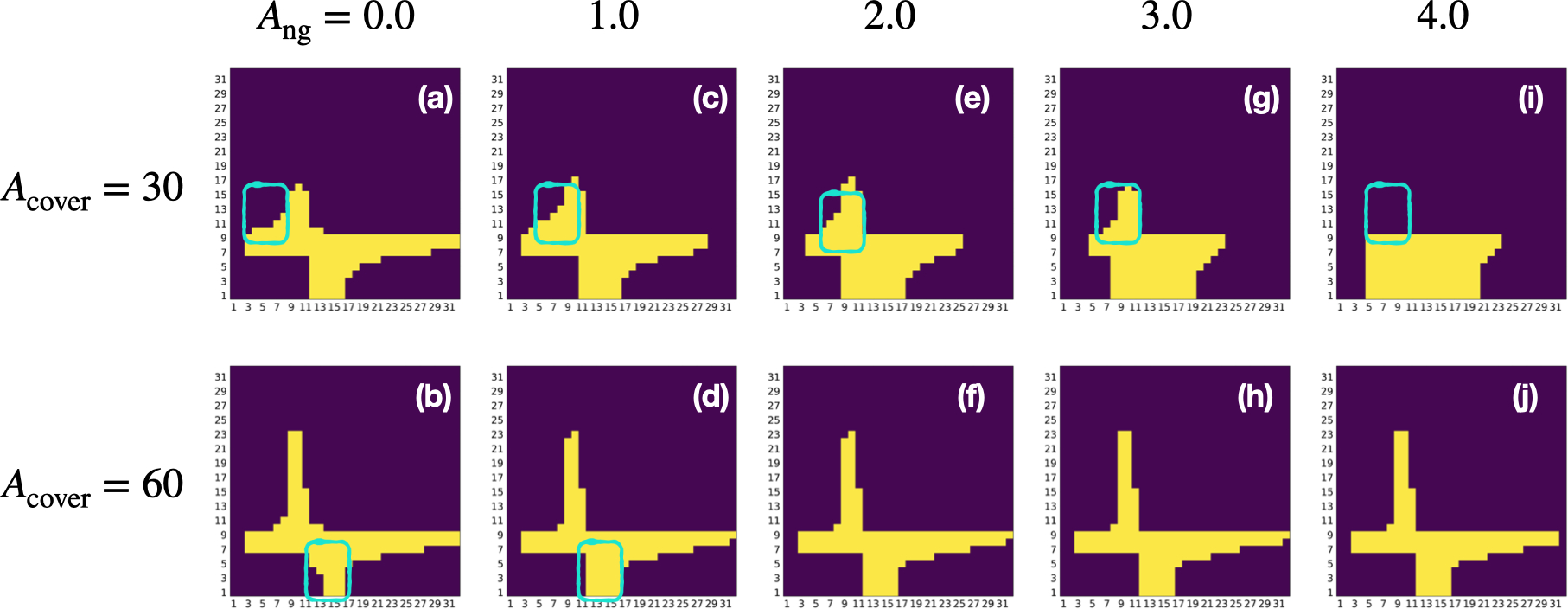}
\caption{Example computation: How the shape of discrete geofences changes when varying $A_\mathrm{ng}$ with different primary weight $A_\mathrm{cover}\in\{30,60\}$. The effect of adjusting the details of boundaries, which are highlighted by solid lines.}
\label{fig-results2-Ang}
\end{figure}

\begin{figure}[htb]
\centering
\subcaptionbox{$10\%$\label{fa1}}[0.16\linewidth]{
    \includegraphics[width=\linewidth]{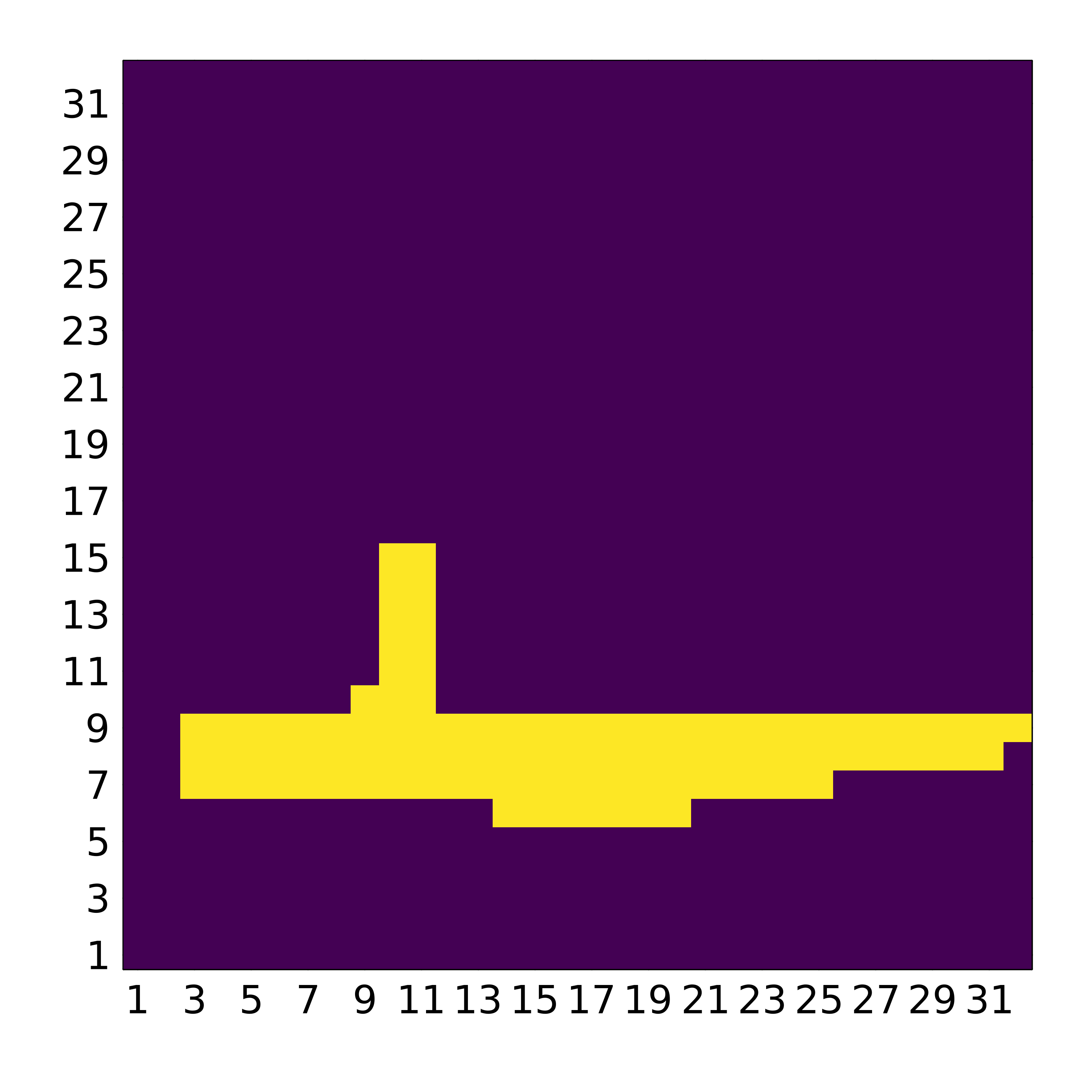}
}
\subcaptionbox{$13\%$\label{fa2}}[0.16\linewidth]{
    \includegraphics[width=\linewidth]{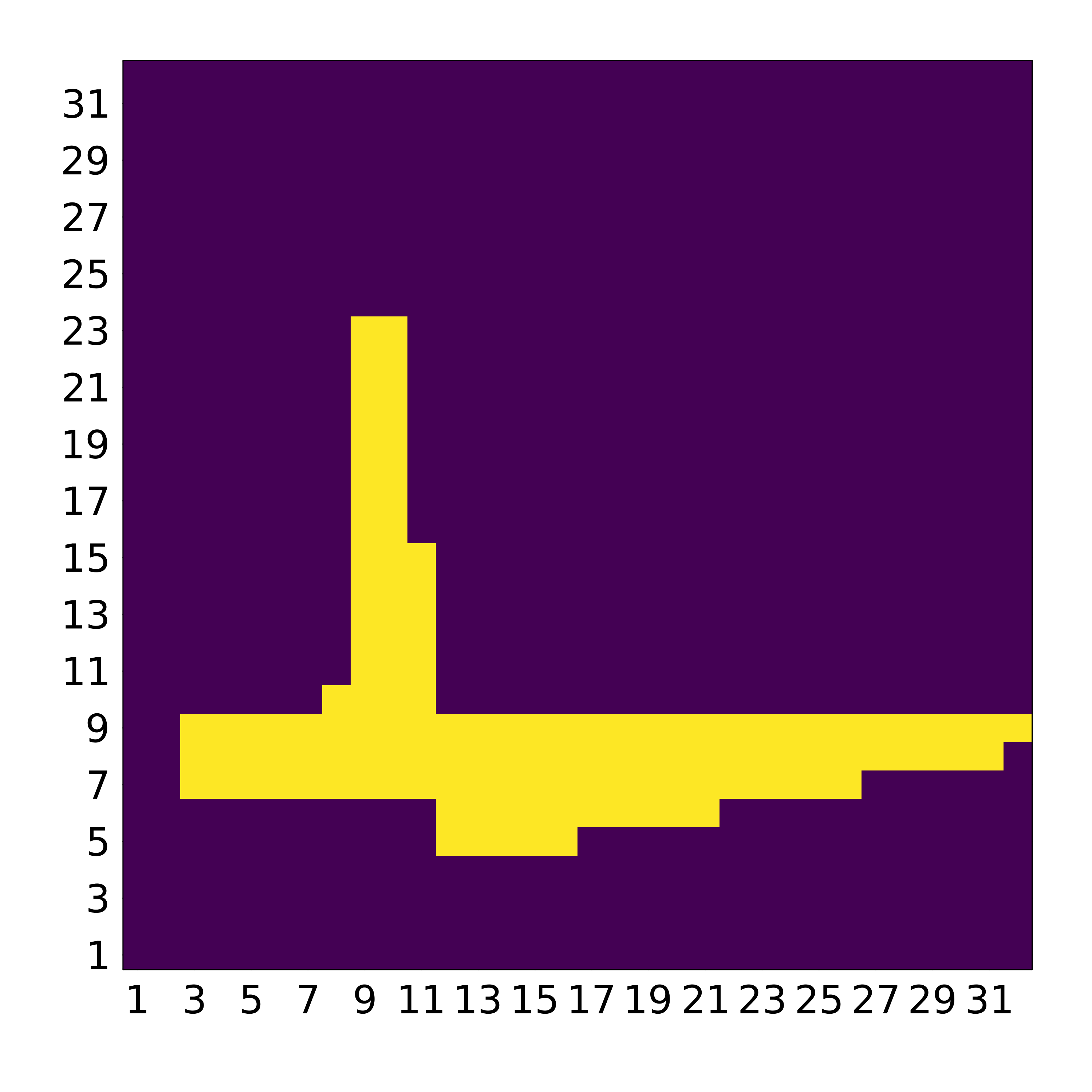}
}
\subcaptionbox{$16\%$\label{fa3}}[0.16\linewidth]{
    \includegraphics[width=\linewidth]{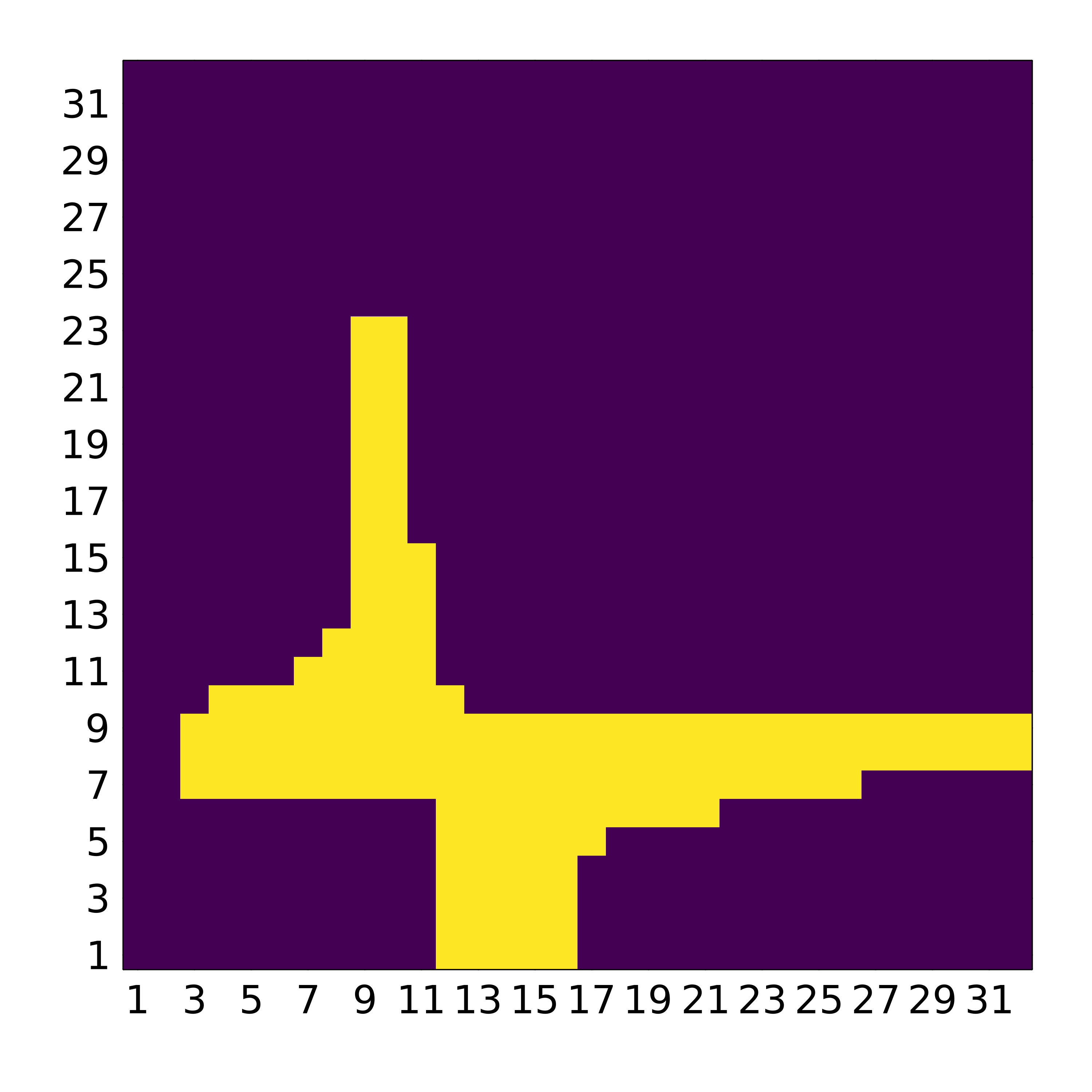}
}
\subcaptionbox{$19\%$\label{fa4}}[0.16\linewidth]{
    \includegraphics[width=\linewidth]{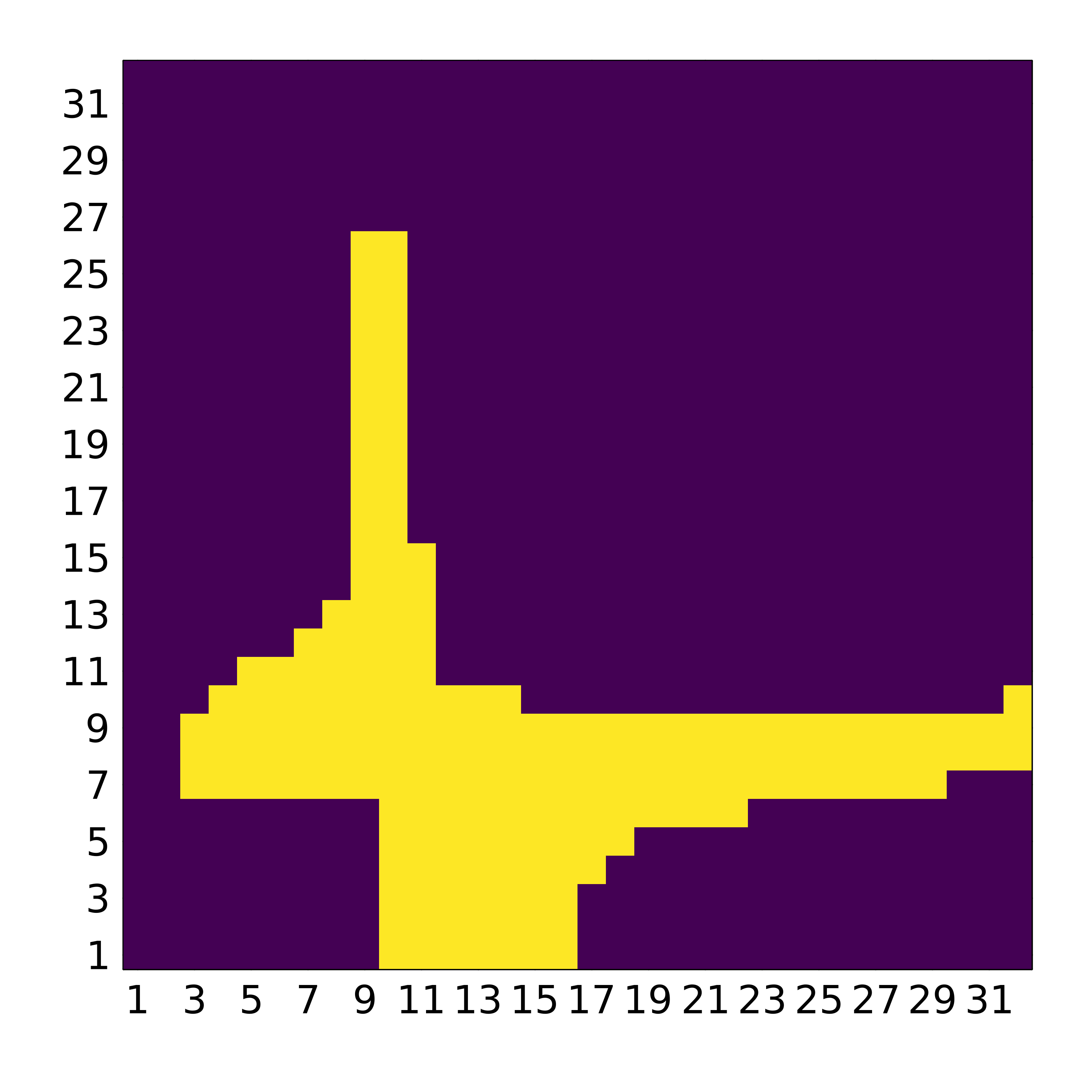}
}
\subcaptionbox{$22\%$\label{fa5}}[0.16\linewidth]{
    \includegraphics[width=\linewidth]{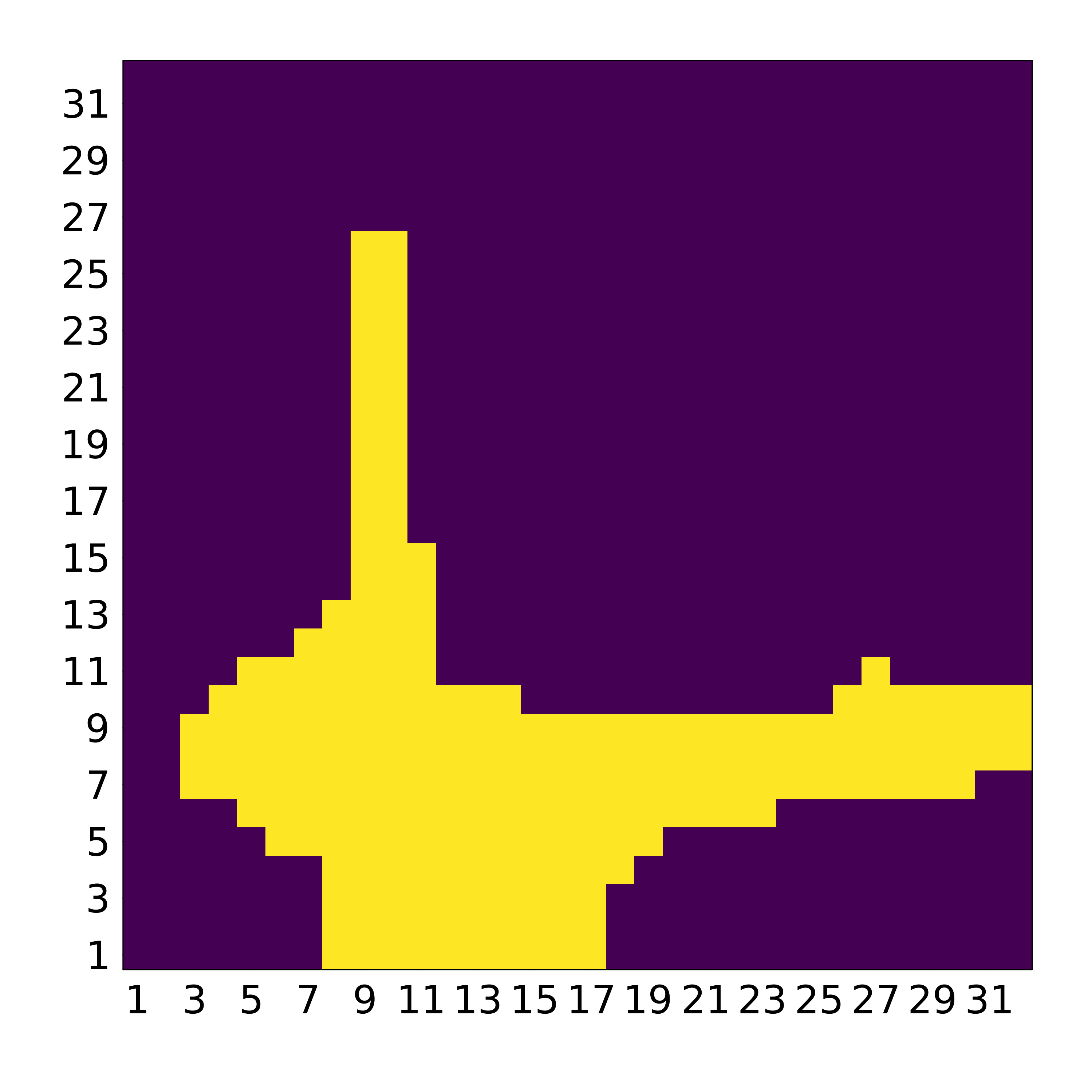}
}
\subcaptionbox{$25\%$\label{fa6}}[0.16\linewidth]{
    \includegraphics[width=\linewidth]{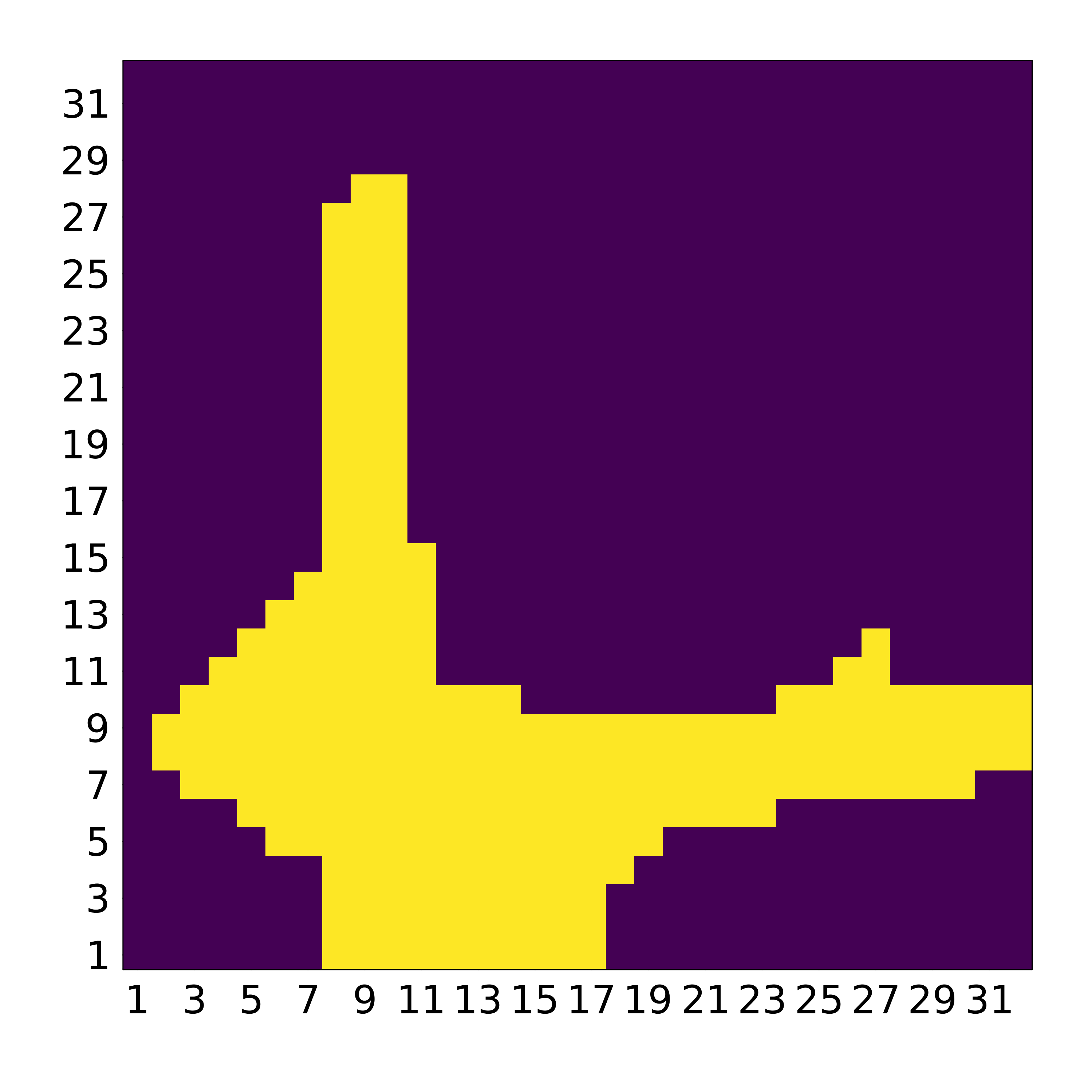}
}
\caption{Example: Variations in $A^\mathrm{max}_\mathrm{area}$ with setting $A^\mathrm{min}_\mathrm{area} := 0.8 A^\mathrm{max}_\mathrm{area}$. Note that $16\%$ means $A_\mathrm{area}^\mathrm{max}=0.16 L^2$.}
\label{fig-results2-Aa}
\end{figure}

The results obtained by varying selected parameters are presented in Figures~\ref{fig-results2-A2D}, \ref{fig-results2-Ang}, and~\ref{fig-results2-Aa}.

In Figure~\ref{fig-results2-A2D}, up to 15\% selected cells and fixed $A_\mathrm{ng}=1.0$, we varied $A_\mathrm{2DW}$ for each primary coefficient $A_\mathrm{cover}\in\{30,60\}$ and computed geofences.
The term $f_\mathrm{2DW}$ depends solely on variables $X$ that represent the geofence shape through the Domain Wall constraints.
Its effect is evident when comparing results for $A_\mathrm{2DW} \in \{0.0, 1.0, 2.0, 3.0, 4.0\}$.
When $A_\mathrm{2DW}=0.0$, other fixed parameters control the selected region, as observed in (a) and (b).
For $A_\mathrm{2DW}>0$, depending on $A_\mathrm{cover}$ and input $V$, the Domain Wall constraints control the shape, as compared in (c) and (d).
For $A_\mathrm{cover}=30$ (the top row of Figure~\ref{fig-results2-A2D}), the increment of $A_\mathrm{2DW}$ prioritizes the square-like island (as in (e), (g), and (i)), rather than including cells having higher values in $V$.
In contrast, for $A_\mathrm{cover}=60$ (the bottom row of Figure~\ref{fig-results2-A2D}), the computed geofences have different shapes (cf. (e) and (f)).
By adjusting the value of $A_\mathrm{2DW}$ with a fixed $A_\mathrm{cover}$, we can obtain discrete geofences with more uniform shapes from a global perspective.

Next, we fix $A_\mathrm{2DW}=1.0$ and vary $A_\mathrm{ng}\in\{0.0,1.0,2.0,3.0,4.0\}$ in contrast to Figure~\ref{fig-results2-A2D}, and the computed geofences are illustrated in Figure~\ref{fig-results2-Ang}.
The term $f_\mathrm{ng}(X)$ depends on both cell-level coverage $V_{i,j}$ represented by the data and geometric shape $X$.
Thus, the algorithm selects whether to include or exclude cells while maintaining the desired shape.
Comparing the results for $A_\mathrm{cover}=30$ and $A_\mathrm{cover}=60$, coefficient $A_\mathrm{ng}$ affects the shape locally.
For $A_\mathrm{cover}=30$ (the upper row in Figure~\ref{fig-results2-Ang}), the shape of the geofence is gradually shifted to a rectangular-like shape~(see (a), (c), (e), and (g)), and finally, (i) gets the almost rectangular-like shape.
For $A_\mathrm{cover}=60$ (the lower row in Figure~\ref{fig-results2-Ang}), the smoothness of the boundary shape slightly changes (see in (b) and (d)), and the entire shape is controlled together with $A_\mathrm{cover}$ (cf. (i) and (j)).
In summary, while $A_\mathrm{ng}$ controls the boundary smoothness of a geofence, it affects locally in contrast to the effect of increasing $A_\mathrm{cover}$; thus, we can set $A_\mathrm{cover}$ first to cover input trajectory data and adjust the shape by modifying $A_\mathrm{ng}$.

Following previous results on $A_\mathrm{2DW}$ and $A_\mathrm{ng}$ with different $A_\mathrm{cover}$, coverage weight $A_\mathrm{cover}$ directly affects the shape according to the input data, and two soft constraints could adjust the details of boundaries, which are highlighted by solid lines in Figures~\ref{fig-results2-A2D} and \ref{fig-results2-Ang}.
From this perspective, the hard constraint in $f_\mathrm{area}(X)$, as shown in Eq.\eqref{eq-obj-total-hard}, should be evaluated as it affects the global shape because it can constrain the number of cells included in the discrete geofence.
Figure~\ref{fig-results2-Aa} shows how the geofence size changes when varying $A^\mathrm{max}_\mathrm{area}$ from 10\% to 25\% of the total cell count ($L^2 = 2^{d} \times 2^{d}$), increasing by 3\% at each step.

In the optimized results, we confirm that the hard constraint effectively controls the degree of coverage on the input trajectory data (cf. Figure~\ref{fig-input_data}).
Once the upper bound is set to be small in (a), the geofence preferentially covers the dense area, enabling us to match human mobility data.
If we increase the size, geofences start to cover the branching trajectories (e.g., top in (b) and bottom in (c)).
If we continue more, geofences can form rectangular-like shapes determined by other parameters (e.g., (d)--(f)).
These computational results confirm that the discrete geofence design allows for generating complex geofence shapes beyond the limitations of the conventional formulation based on a circular shape.

\paragraph{Computational Time}
\label{para-exp1-times}

\begin{figure}[htb]
\centering
\includegraphics[width=110mm]{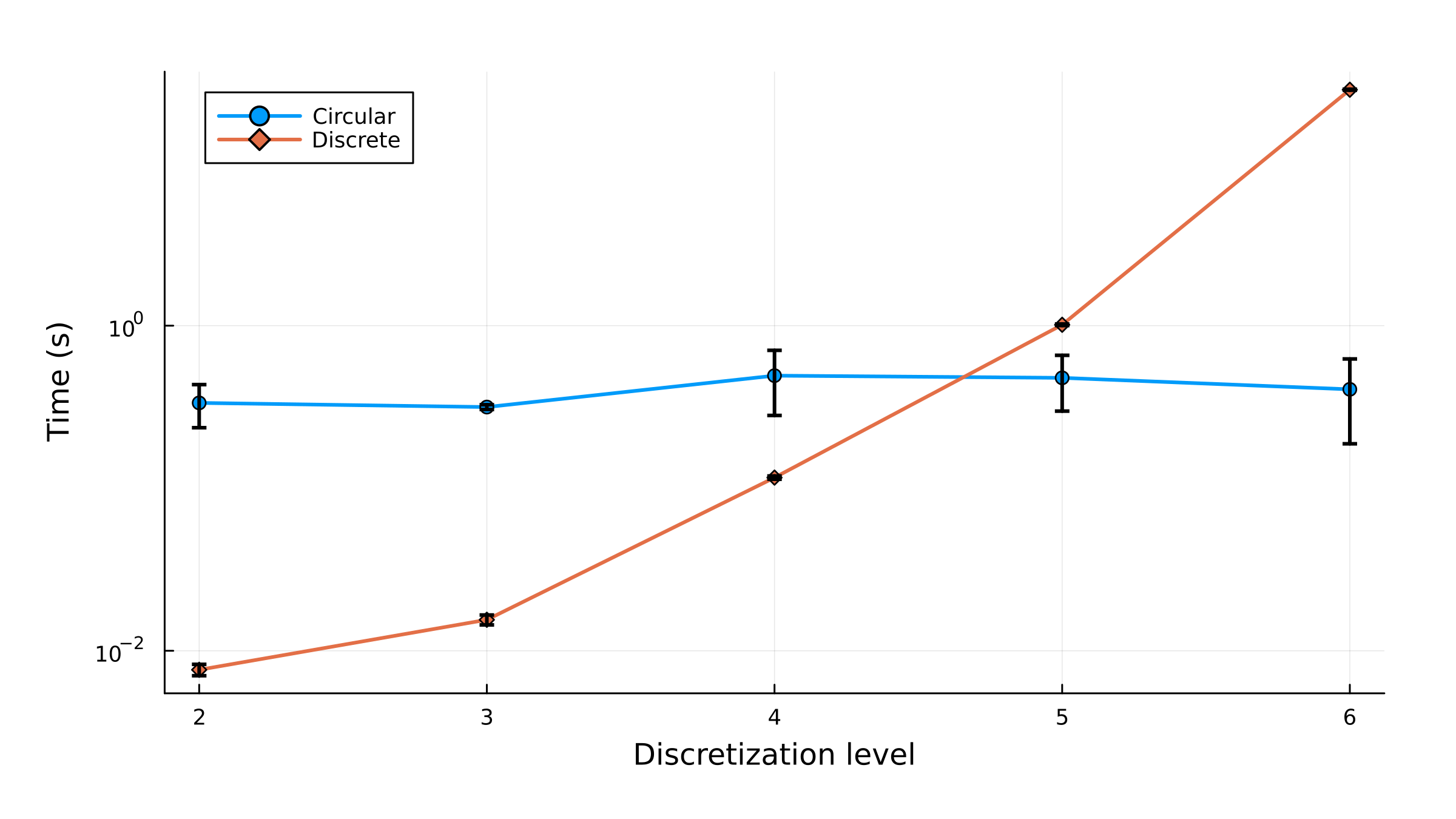}
\caption{Computational time comparison with log-scale time axis: Results for varying resolutions $d\in [2, 3, 4, 5, 6]$ versus the time required for computing a single circular geofence.}
\label{fig-results1-time}
\end{figure}

To investigate the computational performance, we vary discretization resolution $d$ over $[2, 3, 4, 5, 6]$ for fixed parameters, and plot the resulting computational times in Figure~\ref{fig-results1-time}.
For comparison with a reference case, we also plot the times when computing a single circular geofence using a meta-heuristic solver.
For the circular geofence, the default parameter setting stabilized the computational times.
Owing to the curse of dimensionality in the discrete variable matrix, $X$, obtaining an exact solution (when possible) through mathematical optimization solvers becomes computationally intensive at higher resolutions.
Therefore, alternative approaches such as employing heuristic solvers like annealing solvers or adopting a hierarchical strategy could be promising, where candidate geofences are first identified at a coarser discretization level $d$, followed by optimization at finer resolutions for locations where $X_{i,j}=1$.

\paragraph{Coverage Analysis}
\label{para-exp1-coverage}

Because our primary objective function incorporates the term $f_\mathrm{cover}(\GF{i})$ rather than the sum of terms $f(\GF{i})$ and $g^\mathrm{mincover}(\GF{i})$ in circular geofences, we anticipate both improved shape optimization and enhanced user coverage.
To evaluate the actual extent to which geofences cover user movements, we measure the following scores.

\begin{itemize}
\item User-level coverage ratio (UCR, equivalent to Eq.~\eqref{eq-cgf-cover}): The ratio of users that are covered by the computed geofences to the number of all users.
\item User-point-level coverage ratio per user (UPCR): $\mathrm{UPCR}$ is computed by the user-averaged coverage ratios, that is, $\mathrm{UPCR}=\frac{1}{|\mathrm{User}|} \sum_{u\in\mathrm{User}}\frac{\text{\# of user trajectory points inside geofence of }u}{\text{total \# of trajectory points of }u}$, where $\mathrm{User}$ is the set of users in $\DB$.
\end{itemize}

\noindent
Notably, the improvement in UPCR is particularly relevant for applications such as navigation and traffic management, where continuous coverage of user trajectories is essential.

\begin{figure}[htb]
\centering
\includegraphics[width=0.19\linewidth]{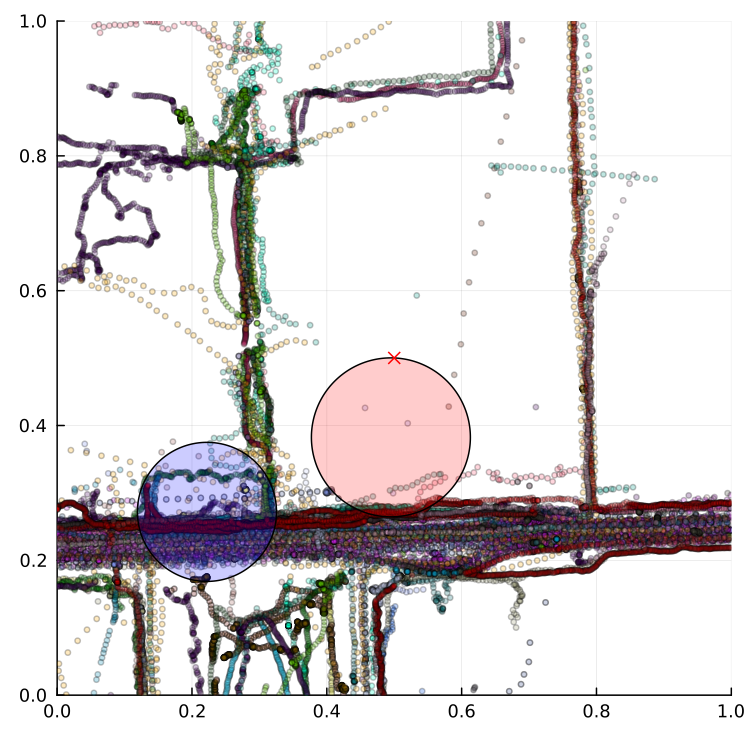}
\includegraphics[width=0.75\linewidth]{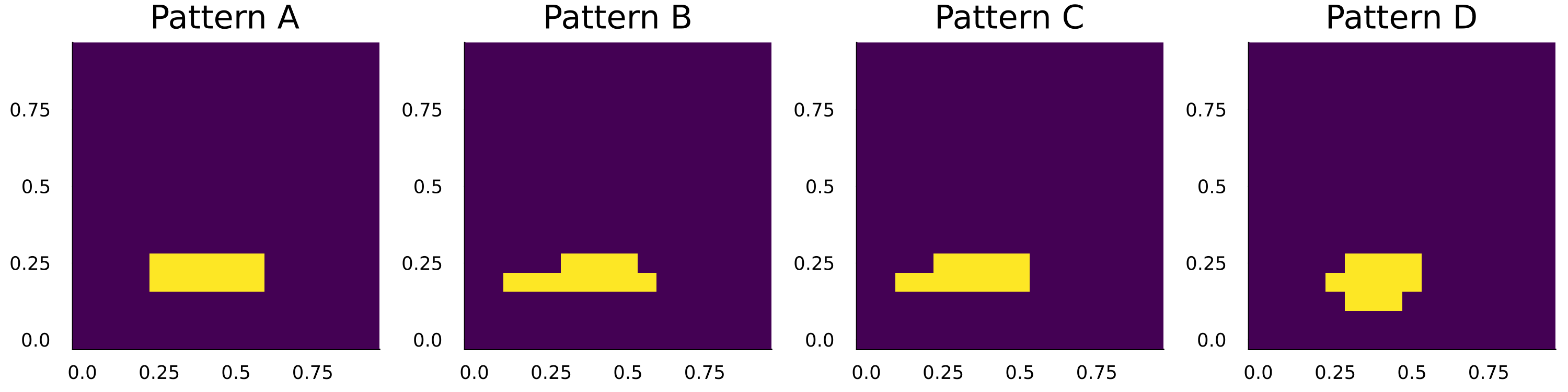}
\caption{Two circular geofences (red: original, blue: cover-oriented) and distinct discrete geofence patterns (A, B, C, and D) in the coverage experiment.}
\label{fig-coverage-pattern}
\end{figure}

To evaluate these scores, we varied parameters as $A_\mathrm{cover}\in\{30,60\}, A_\mathrm{2DW}\in\{0.0, 1.0, 3.0\}, A_\mathrm{ng}\in\{0.0, 1.0, 3.0\}$; from all their combination, we computed four distinct discrete geofence shapes.
For CGFs, we call by \textit{CGF~(original)} the original formulation that maximizes $f(\GF{i}) + g^\mathrm{mincover}(\GF{i})$.
In addition, we implement \textit{CGF~(cover-oriented)} that mimics our covering-oriented discretized version as follows: put the primary objective function as $g^\mathrm{mincover}(\GF{i})$ under a constraint in which $R$ is optimized within $[R^\star-\epsilon, R^\star+\epsilon]$, where $R^\star$ is the radius of the solution of the original CGF and $\epsilon=0.01$ to keep the original size.
The resulting two circular and four distinct discrete geofences using the above parameters are illustrated in Figure~\ref{fig-coverage-pattern}, named Patterns A, B, C, and D.

We then compare two CGFs and four patterns of discrete geofences from the coverage viewpoint.
For those geofences, UCR and UPCR scores are evaluated and summarized in Table~\ref{table-ucr}.
We confirm that discrete geofences generate more cover-oriented shapes from both UCP and UPCR viewpoints, as these scores are larger than those of the original and cover-oriented CGFs.

\begin{table}[htb]
\caption{Evaluated UCR and UPCR scores. The underline score in Circular (Cover-oriented) means that only this result was measured by repeatedly solving problems to consider the randomness in the meta-heuristic solver.}
\label{table-ucr}
\centering
\begin{tabular}{l|cc|cccc}
\toprule
 & \multicolumn{2}{c}{Circular} & \multicolumn{4}{c}{Proposed method} \\
 & Original & Cover-oriented & Pattern A & Pattern B & Pattern C & Pattern D \\
\midrule
UCR & $0.531$ & \underline{$0.775~(0.159)$} & $0.969$ & $0.969$ & $0.969$ & $0.969$ \\
\midrule
UPCR (mean) & $0.013$ & $0.103$ & $0.246$ & $0.253$ & $0.260$ & $0.223$ \\
UPCR (std) & $(0.024)$ & $(0.064)$ & $(0.125)$ & $(0.142)$ & $(0.127)$ & $(0.128)$ \\
\bottomrule
\end{tabular}
\end{table}

\begin{figure}[htb]
\centering
\subcaptionbox{UPCR by original CGFs versus proposed method.\label{fig-coverage-org}}[0.32\linewidth]{
    \includegraphics[width=\linewidth]{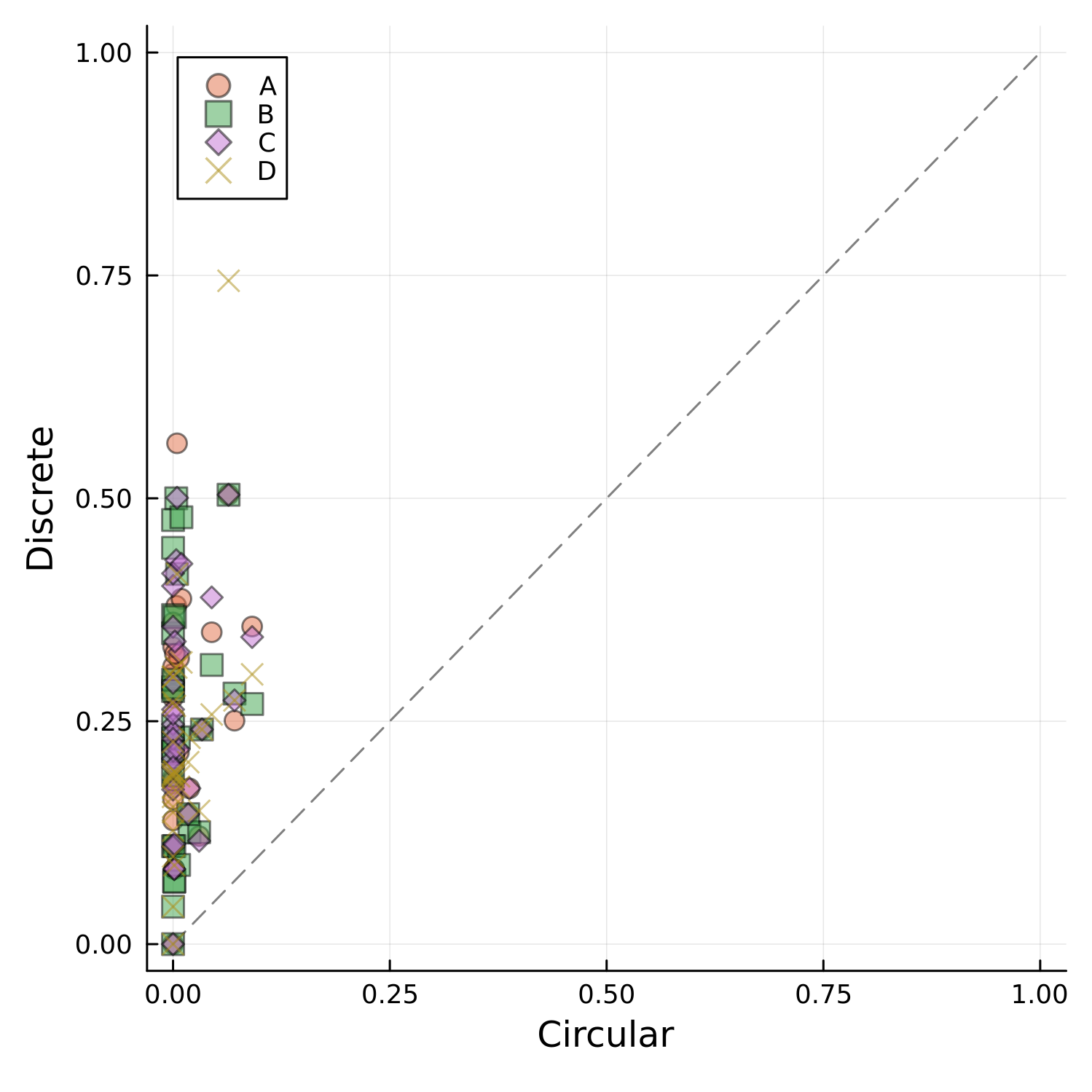}
}
\hspace{2em}
\subcaptionbox{UPCR by cover-oriented CGFs versus proposed method.\label{fig-coverage-oriented}}[0.32\linewidth]{
    \includegraphics[width=\linewidth]{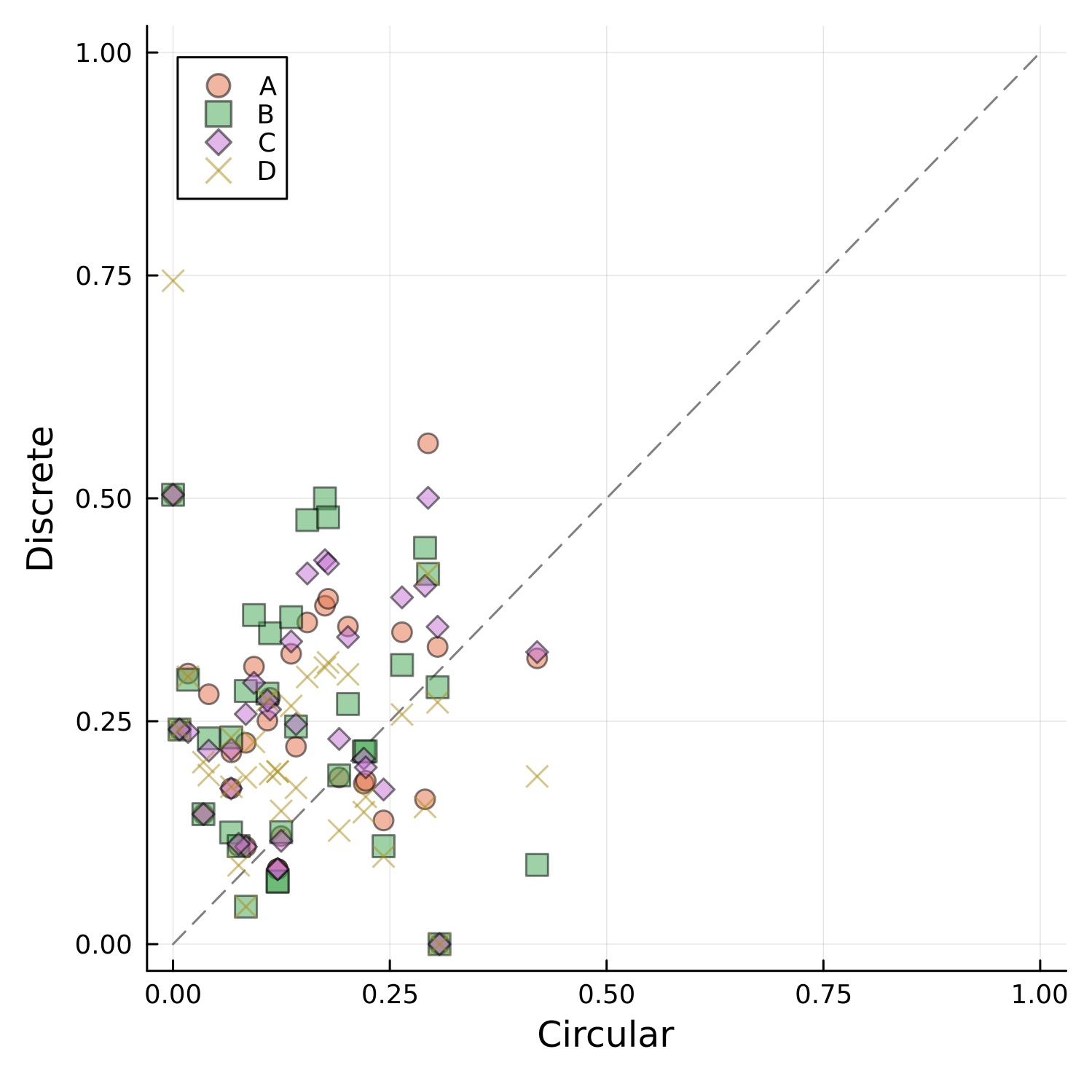}
}

\caption{Coverage analyses: Four discrete geofences and (a) original CGF and (b) cover-oriented CGF, where scatter visualization illustrates the method with higher user-level coverage. Scatter point markers (A, B, C, and D) in different colors correspond to patterns A, B, C, and D, respectively.}
\end{figure}

To compare user-level results, we also illustrate scatter plots per user (i.e., a scatter point corresponds to a user in set $\mathrm{User}$), in Figures~\ref{fig-coverage-org} and~\ref{fig-coverage-oriented}, respectively.
In the figures, the $x$-axis and $y$-axis represent circular and discrete geofences, respectively.
The observation that points lie above the diagonal line indicates that discrete geofences are more effective in terms of user-point-level coverage.
These results also indicate that the discrete approach shows better trajectory coverage than circular approaches, particularly for user perspectives.
Even if the original formulation is modified to maximize the coverage, the result demonstrates the effectiveness and representation capability using discrete shapes.

\subsubsection*{Synthetic Data Case}
\label{subsub-results-dgf-synthetic}

\begin{figure}[htb]
\centering

\subcaptionbox{Data 1}[0.48\linewidth]{
    \includegraphics[width=0.5\linewidth]{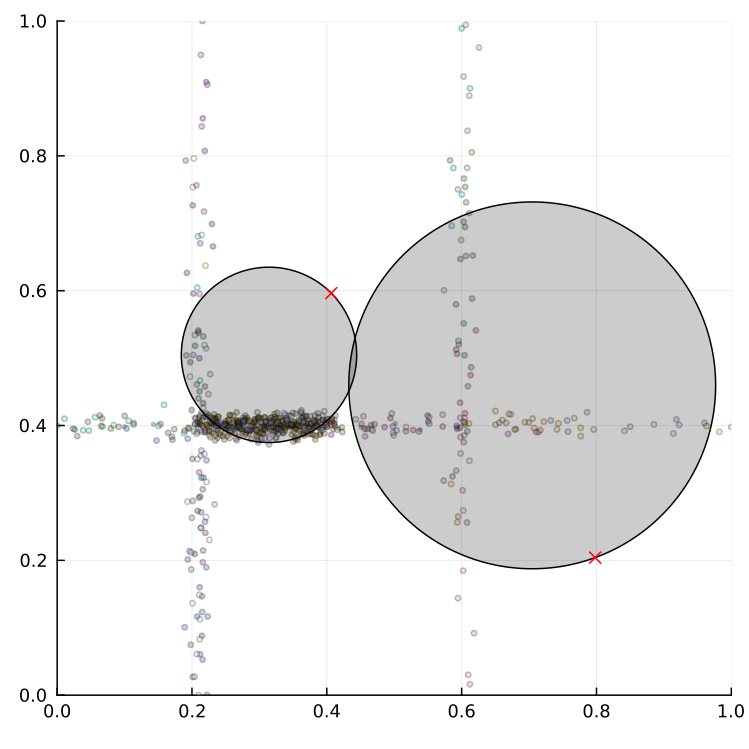}
}
\subcaptionbox{Data 2}[0.48\linewidth]{
    \includegraphics[width=0.5\linewidth]{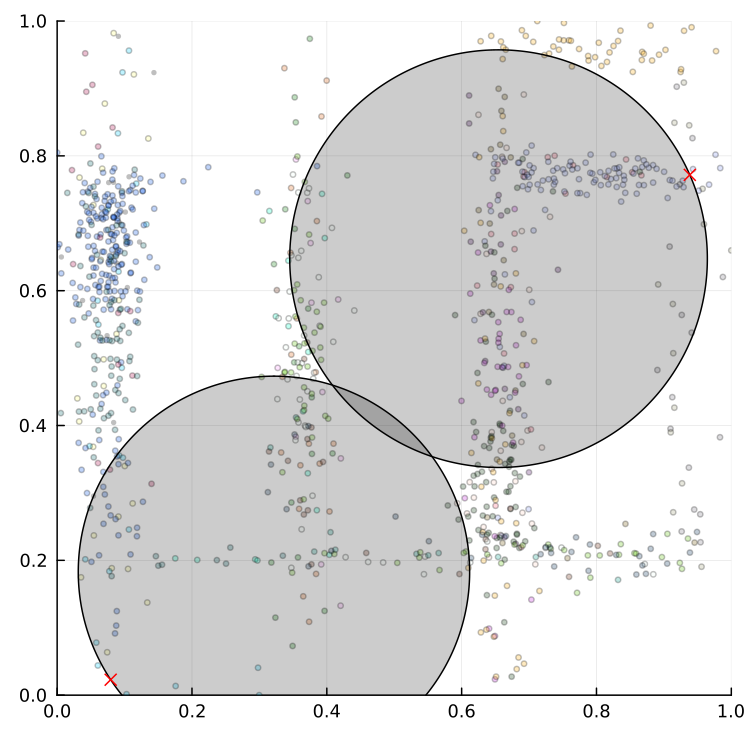}
}

\caption{Examples of high-density environments using synthetic data (Data 1 and Data 2), illustrated together with circular geofences.}
\label{fig-results3-inputs}
\end{figure}

Using synthetic data, we examined computational results in a high-density environment with multiple POIs beyond the single POI case in real-world data.
For evaluation, we present two types of example data, Data~1 and Data~2, which have two POIs~(a and b), and optimize them using a circular geofence, as shown in Figure~\ref{fig-results3-inputs}.
The optimization settings are similar to those in the previous section.
The optimized circular geofences corresponding to two POIs each form the smallest possible circles that encompass the minimum required coverage area.
These cases demonstrate the potential for notification zones to overlap with nearby POIs.

\begin{figure}[htb]
\centering

\subcaptionbox{a: 15\%}[0.16\linewidth]{
    \includegraphics[width=0.9\linewidth]{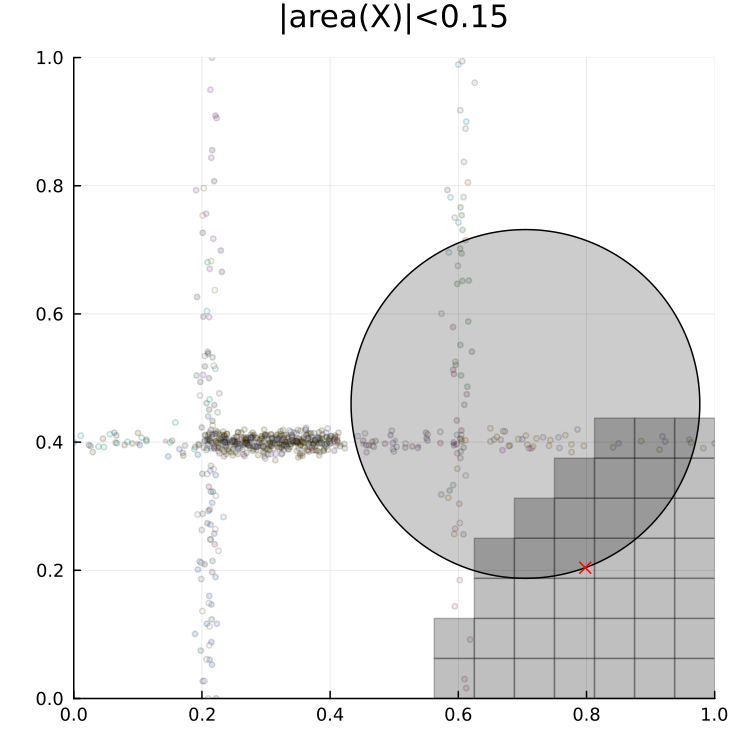}
}
\subcaptionbox{a: 20\%}[0.16\linewidth]{
    \includegraphics[width=0.9\linewidth]{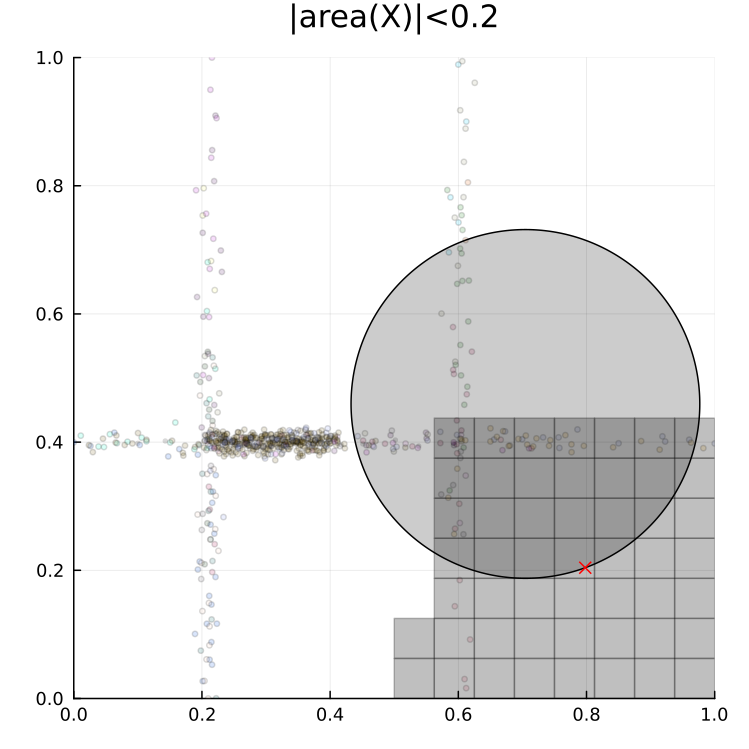}
}
\subcaptionbox{a: 25\%}[0.16\linewidth]{
    \includegraphics[width=0.9\linewidth]{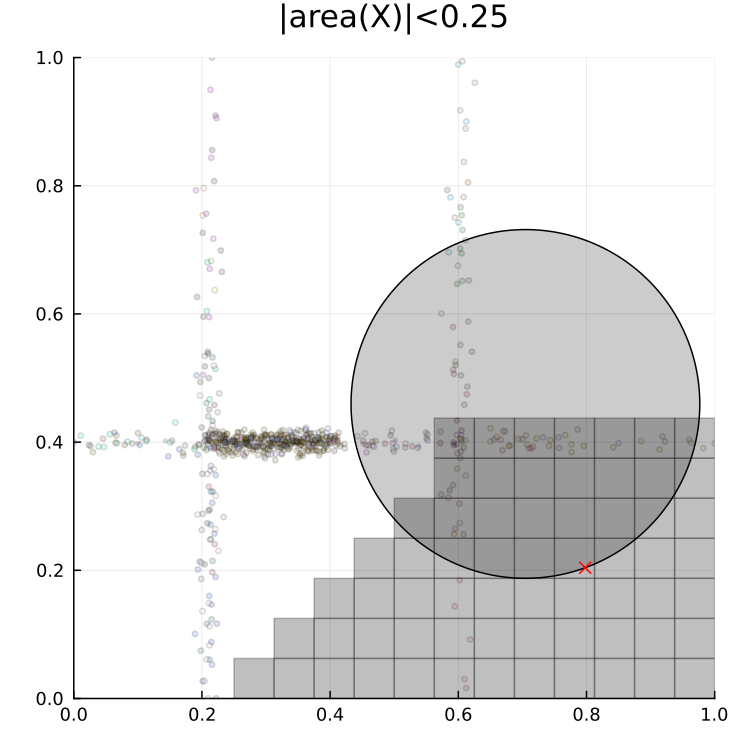}
}
\subcaptionbox{b: 15\%}[0.16\linewidth]{
    \includegraphics[width=0.9\linewidth]{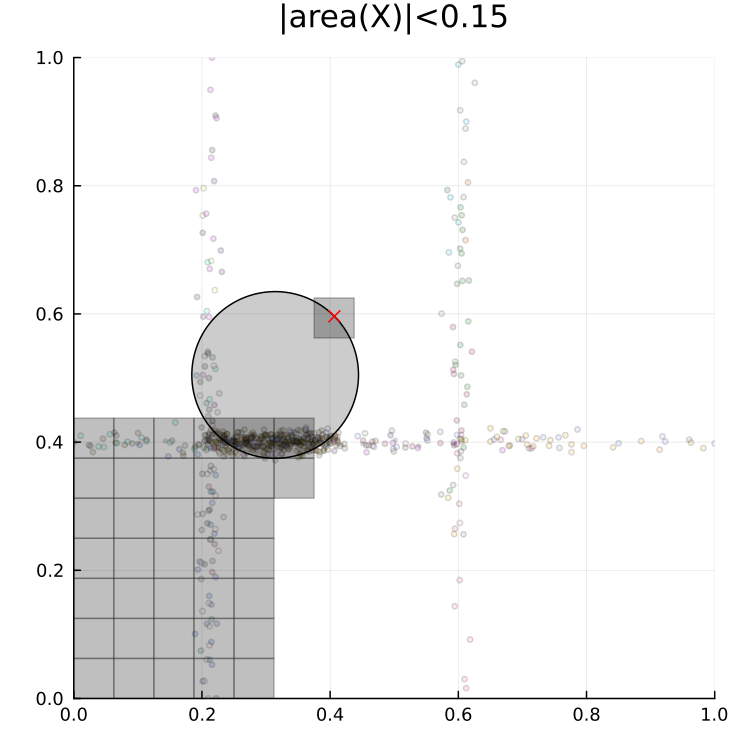}
}
\subcaptionbox{b: 20\%}[0.16\linewidth]{
    \includegraphics[width=0.9\linewidth]{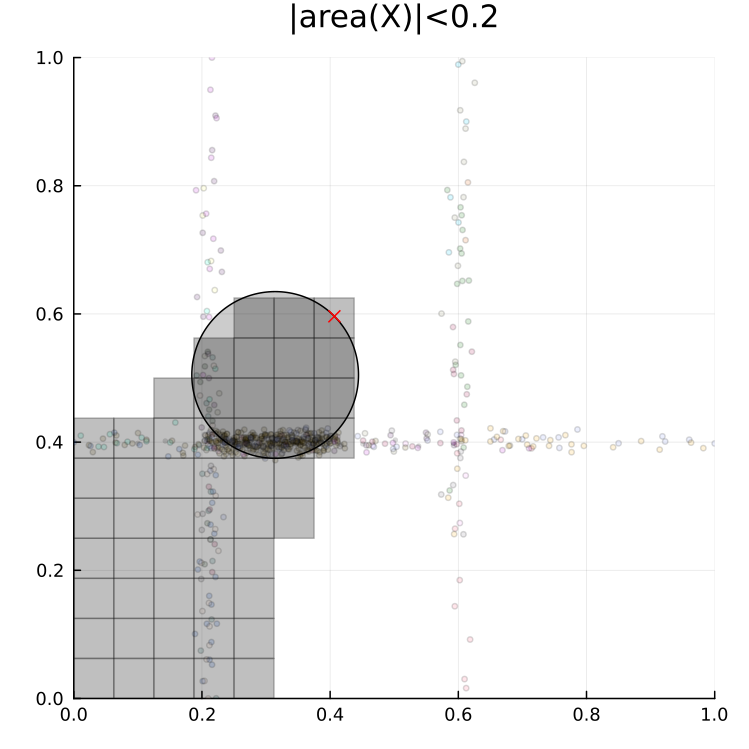}
}
\subcaptionbox{b: 25\%}[0.16\linewidth]{
    \includegraphics[width=0.9\linewidth]{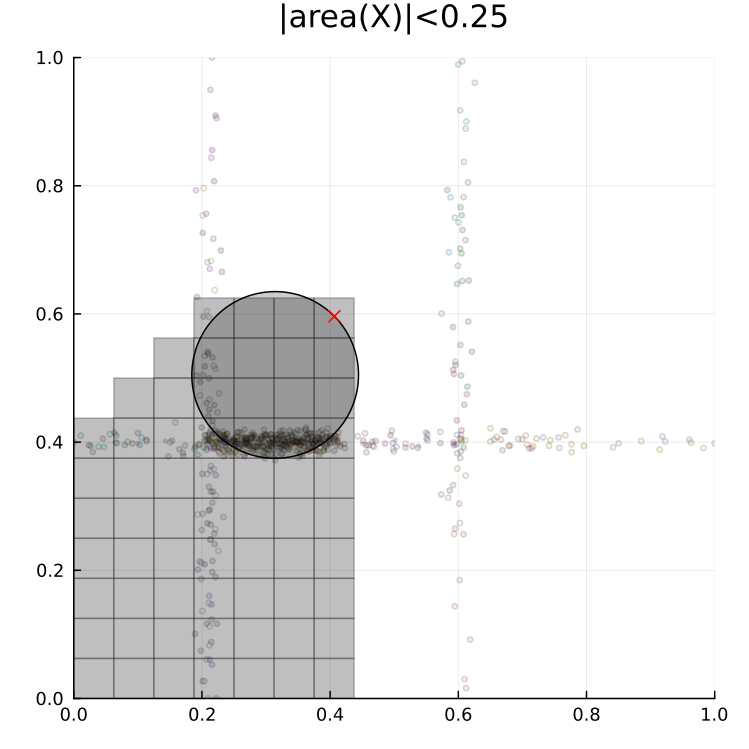}
}
\caption{Examples of discrete geofences generated for POIs a and b in Data 1 with increasing area.}
\label{fig-results3-data1}
\end{figure}

\begin{figure}[htb]
\centering

\subcaptionbox{a: 15\%}[0.16\linewidth]{
    \includegraphics[width=0.9\linewidth]{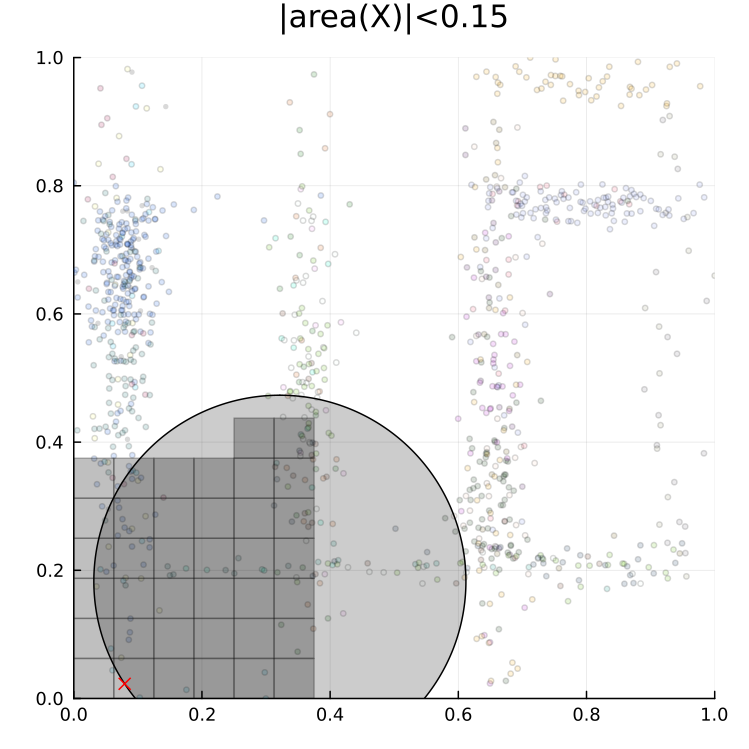}
}
\subcaptionbox{a: 20\%}[0.16\linewidth]{
    \includegraphics[width=0.9\linewidth]{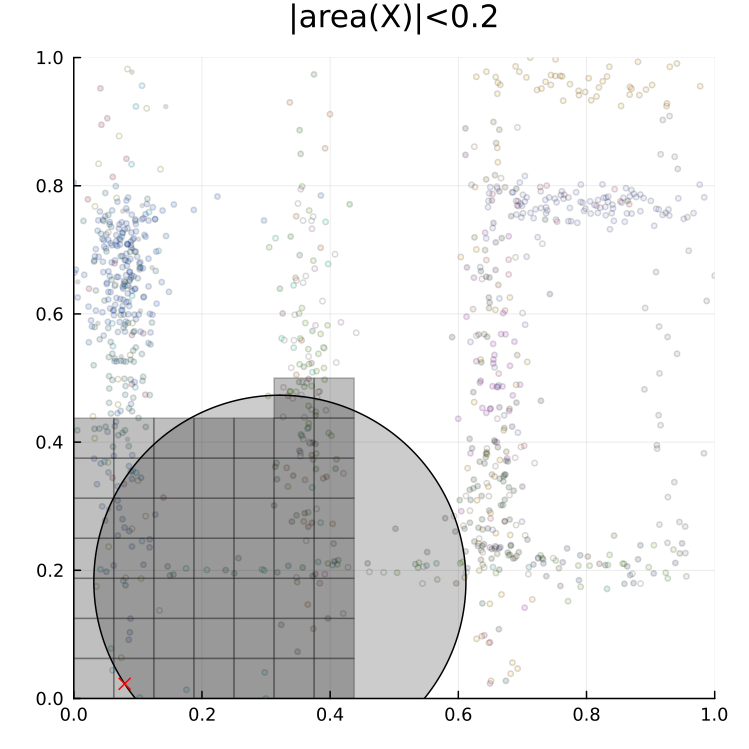}
}
\subcaptionbox{a: 25\%}[0.16\linewidth]{
    \includegraphics[width=0.9\linewidth]{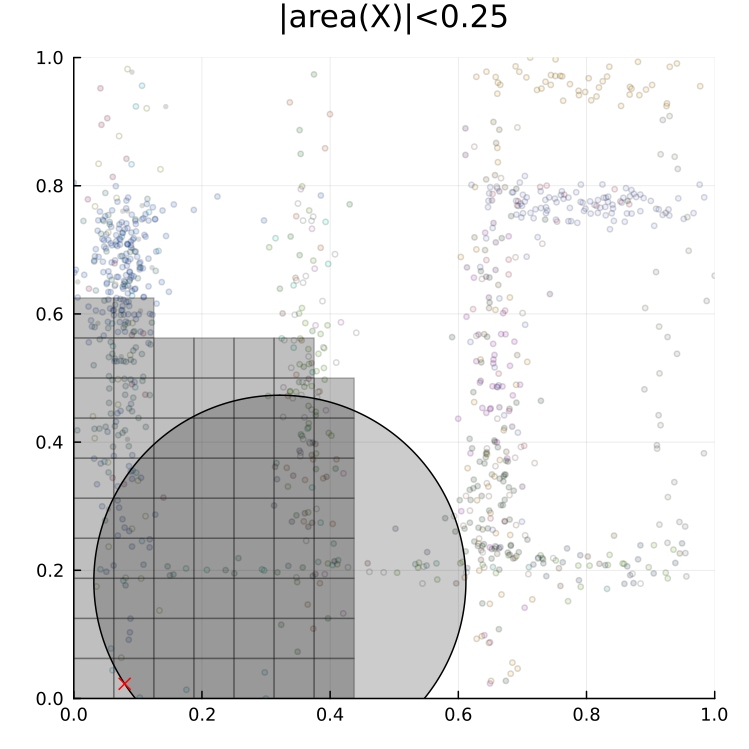}
}
\subcaptionbox{b: 15\%}[0.16\linewidth]{
    \includegraphics[width=0.9\linewidth]{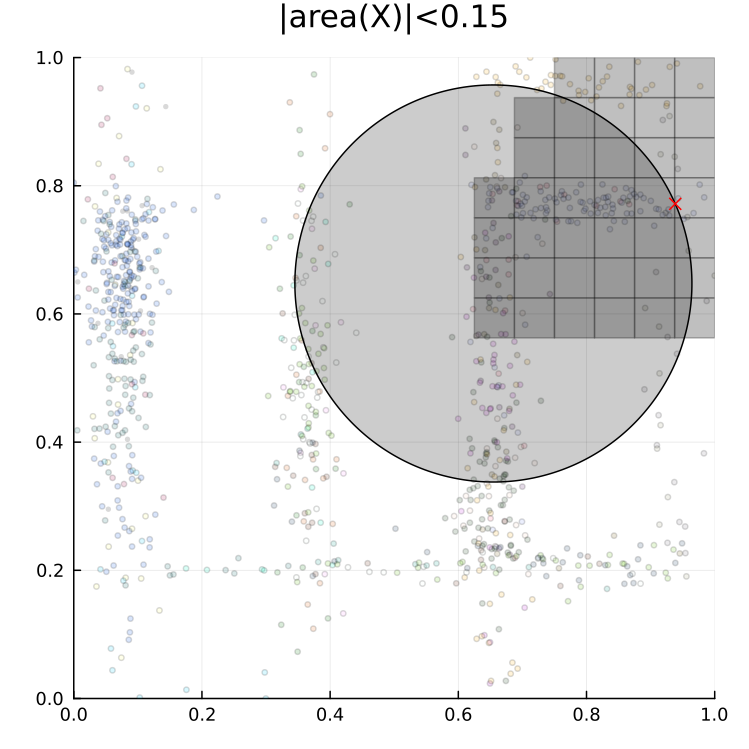}
}
\subcaptionbox{b: 20\%\label{Data2-l1}}[0.16\linewidth]{
    \includegraphics[width=0.9\linewidth]{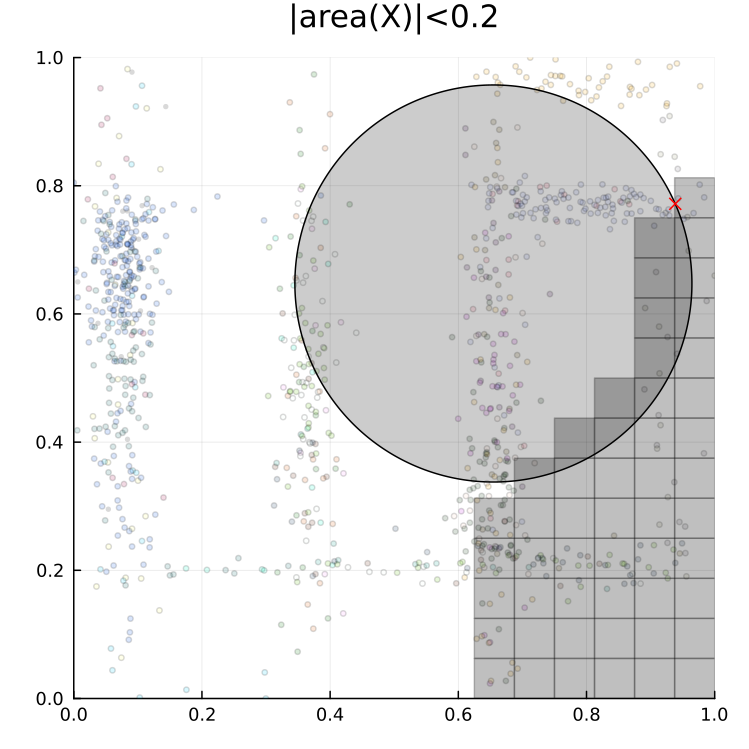}
}
\subcaptionbox{b: 25\%\label{Data2-l2}}[0.16\linewidth]{
    \includegraphics[width=0.9\linewidth]{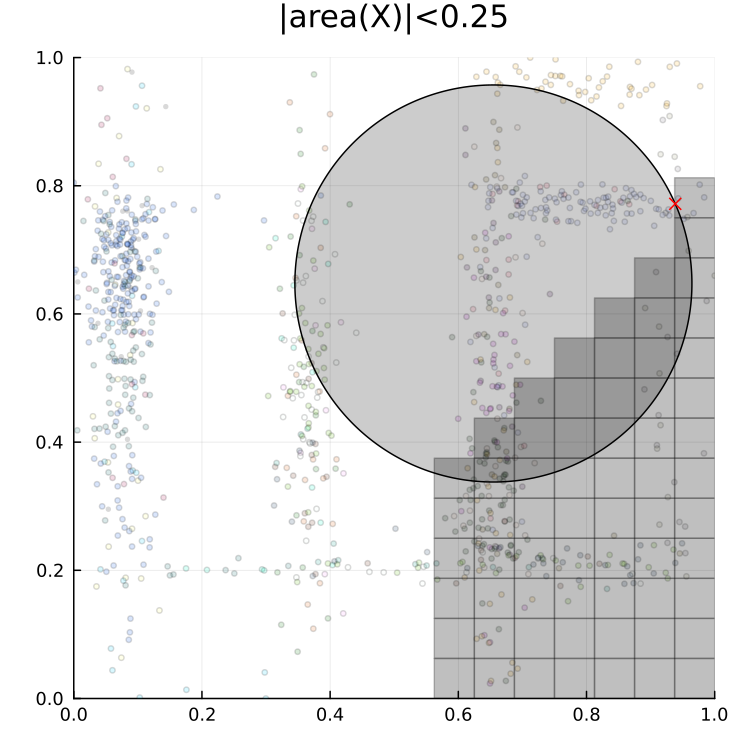}
}
\caption{Examples of discrete geofences generated for POIs a and b in Data 2 with increasing area.}
\label{fig-results3-data2}
\end{figure}

We then discretize these data using a $d=4$ resolution and compute the discrete geofences, which are overlaid on the circular geofences for visualization.
The computational results with $A_\mathrm{cover}=10, A_\mathrm{ng}=1.0, A_\mathrm{2DW}=1.0, A_\mathrm{area}^\mathrm{max}=0.15$ for Data 1 are shown in Figure~\ref{fig-results3-data1}, whereas those for Data 2 are shown in Figure~\ref{fig-results3-data2}, each separately for each POI out of POIs a and b.

Figure~\ref{fig-results3-data1} shows the optimized discrete geofences for Data 1.
The optimized geofences become rectangles to progressively cover the areas surrounding POIs a and b while forming maximally connected regions.
Although the coverage definition differs from that in Eq.~\eqref{eq-cgf-mincover}, which specifies minimum coverage, all computed results consistently include points surrounding both POIs where GPS data is observed.
Specifically, for circular shapes, the algorithm selects circular forms that barely touch the data, while it outputs connected regions while ensuring coverage for discrete shapes.
For POI b, because Data 1 contains more data points in its lower left region, the results show similar trends for both circular and discrete cases.
Because the area constraints are relaxed, the coverage area expands from the lower left, ultimately producing a shape that includes the circular geofence.
These results indicate that the optimization problem in Eq.~\eqref{eq-obj-total-hard} exhibits behavior that optimizes the objective function derived from data coverage while simultaneously obtaining connected regions.

Similarly, we interpret the computational results of Data 2 shown in Figure~\ref{fig-results3-data2}.
For POI a in the left section, we observe coverage-dependent regions similar to those observed in Data 1. The exploration behavior demonstrates a more proactive approach to covering data points by varying the search area size.
For POI b in the right section, the behavior resembles that of POI a in Data 1; however, it exhibits coverage patterns distinct from circular shapes. Furthermore, as the search progresses, the solution shape significantly transforms (from a coverage pattern in the upper right to one in the lower right).
These results demonstrate that we can obtain shapes that are more appropriate for the given data by varying the coefficients of the objective function and constraints.
While some data types may yield results that resemble rectangular shapes rather than circles, the distribution of GPS points may require different approaches to obtain contiguous regions compared to circular cases.

\section*{Discussions}
\label{sec-discussions}

\subsection*{Summary of Findings}
\label{subsec-dis-summary}

\paragraph*{Parameter Sensitivity: Shape Smoothness and Connectivity}
\label{para-diss-parameter}

By varying $A_\mathrm{2DW}$ of the connection penalty term for discrete geofences between $0.0$ and $4.0$, we observed suppression of elongated spike-like cells and obtained more rectangular, cohesive structures, as illustrated in Figure~\ref{fig-results2-A2D}.
This demonstrates that $A_\mathrm{2DW}$ controls the complexity of the optimized geofence shape.

Similarly, increasing $A_\mathrm{ng}$ of the smoothness penalty term for neighborhood smoothness between $0.0$ and $4.0$ reduced cell-level local selection, yielding a smoother, more cohesive region, as shown in Figure~\ref{fig-results2-Ang}.
Therefore, $A_\mathrm{ng}$ functions as a local regularization term for creating large connected regions when adjusted in conjunction with $A_\mathrm{2DW}$.

Furthermore, gradually relaxing the upper limit percentage $[A^\mathrm{min}_\mathrm{area}, A^\mathrm{max}_\mathrm{area}]$ of the discrete geofence area (i.e., the number of cells) from 10\% to 20\% expanded the discrete geofences concentrically, revealing a linear trade-off between coverage and area, as depicted in Figure~\ref{fig-results2-Aa}.
Consequently, adjusting the area constraint according to the purpose of covering human mobility data, which could be constrained on geospatial features (e.g., road geometry), is a promising data-driven approach for notifying users in geofence applications.

\paragraph*{Advantages of Weighted Cover Functions}
\label{para-dis-weighted-cover}

The weighted coverage function, $f_\mathrm{cover}(X)$, which is inversely proportional to distance from POIs, effectively discourages selection of distant high-density cells.
This encourages concentration on high-value cells surrounding POIs.
Compared to simple summation-based coverage functions (i.e., Eq.~\eqref{eq-dgf-cover} with $C_{i,j}=1$), the weighted approach with Eq.~\eqref{eq-dgf-weighted-cover} demonstrates better control over excessive expansion, preventing both trivial solutions that select all cells and solutions that merely include POIs without adequately covering them.

\paragraph*{Multiple POI and High-Density Scenario}
\label{para-dis-application}

\begin{figure}[htb]
\centering

\subcaptionbox{Data 1}[0.45\linewidth]{
    \includegraphics[width=0.5\linewidth]{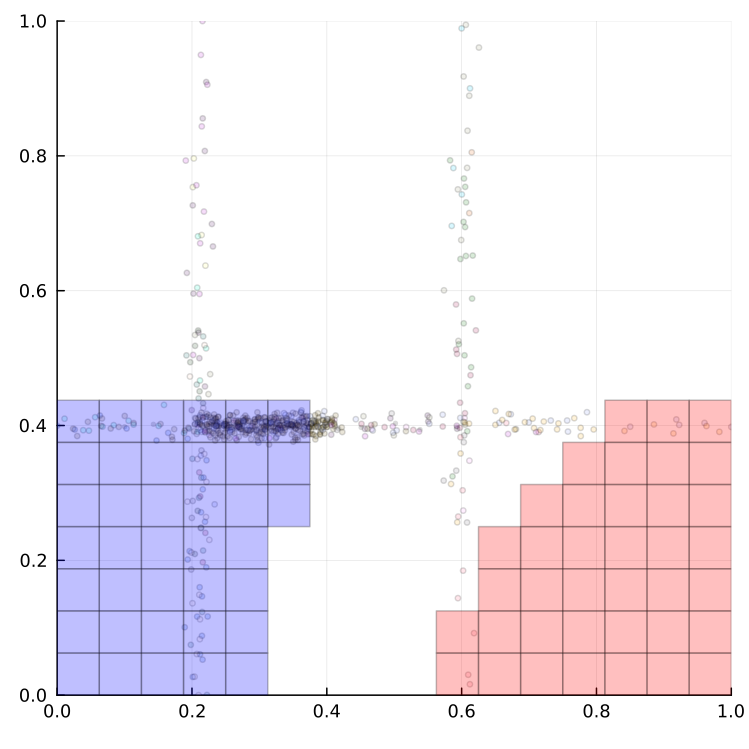}
}
\subcaptionbox{Data 2}[0.45\linewidth]{
    \includegraphics[width=0.5\linewidth]{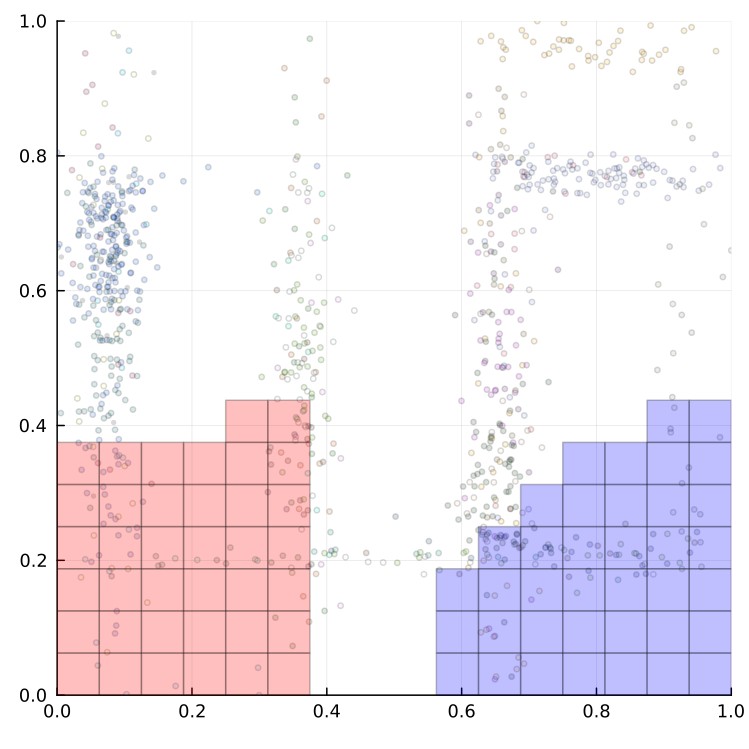}
}
\caption{Overlaid visualizations of two geofences computed for two POIs in Data 1 and Data 2.}
\label{fig-results3-data1-data2}
\end{figure}

In the artificial dataset case study involving a high-density environment with two POIs, we observed that while circular geofences tend to overlap extensively, the proposed discrete geofences tend to extend asymmetrically toward high-value directions for each POI, as illustrated in Figures~\ref{fig-results3-data1} and~\ref{fig-results3-data2}.
This result indicates that the proposed approach successfully maintained high coverage while avoiding overlaps, which are illustrated in Figure~\ref{fig-results3-data1-data2}.
Notably, parameters in these experiments are adjusted to make the two geofences not overlap, resulting in rectangular-like results that do not match the input data.
To adjust the relationship between two geofences, we should carefully control parameters or design the two geofences simultaneously with additional constraints, which could be studied in future work.

For specific data examples, we observed that for POI b in Data 2, relaxing the area constraints dynamically shifted the coverage region from the upper-right to lower-right direction, as shown in Figures~\ref{Data2-l1} and \ref{Data2-l2}.
This result demonstrates the expected adaptive placement behavior of discrete geofences, which dynamically adjust according to human movement trajectories along roads and spatial constraints imposed on geofence coverage.

\subsection*{Practical Design Guideline}
\label{subsec-dis-framework}

Our design approach is to select $A_\mathrm{cover}$ and $[A_\mathrm{min}, A_\mathrm{max}]$ first, then set $A_\mathrm{2DW}$, and finally moderate $A_\mathrm{ng}$ to control the shape of discrete geofences.
Based on our formulations and experimental results presented in this paper, the following design guidelines have been established:
\begin{itemize}
  \item For initial parameters, set $A_\mathrm{area}$ from $10\%$ to $25\%$ of the maximum number of cells. Begin the search with $A_\mathrm{2DW} \approx 1.0$ and $A_\mathrm{ng} \approx 1.0$. Initialize $A_\mathrm{cover}$ according to cell size, data characteristics, and whether coverage is required or not.
  \item For coarsely shaped areas, increase $A_\mathrm{2DW}$ and re-examine the solution.
  \item When coverage is insufficient, relax either $A_\mathrm{cover}$ or $A_\mathrm{area}$ and expand the discrete geofences.
\end{itemize}

\subsection*{Limitations and Future Directions}
\label{subsec-dis-limitation}

While this study independently optimized each POI (as in Figures~\ref{fig-results3-data1} and~\ref{fig-results3-data2}), future work should address global optimization that accounts for interactions between POIs, such as exclusive zones and distance constraints.
Although we provided design guidelines in the previous section, there remains room for further refinement through more precise parameter tuning and data-specific parameterization methods.
Based on our current procedure above, the following extensions can be applicable in future studies.
\begin{itemize}
   \item When handling multiple POIs, optimize each POI independently and then compare overlays to facilitate suppression of overlaps. Alternatively, add hard constraints to explicitly avoid overlaps.
  \item If solutions become diffuse, increase the distance decay parameter, $\alpha$, from $0.5$ to $1.0$ to enable distance-dependent selection.
  \item If the computed solution has discontinuous regions (e.g., holes), post-processing can be applied to clean up solutions.
\end{itemize}

Another important practical consideration is the exponential increase in computational complexity as resolution $d$ increases.
By combining this approach with efficient search methods, such as hierarchical optimization and utilizing high-performance solvers like quantum annealing, we can achieve greater scalability.
Additionally, future extensions should explore incorporating dynamic POIs, time-series demand patterns, and pedestrian reachability considerations into the model.

\section*{Conclusions}
\label{sec-conclusion}

To provide enriched mobility experiences by presenting users with appropriate information, this paper focuses on geofences as a mathematical model that considers physical geographic regions.
First, we investigate the limitations of existing methods and identify two key challenges, particularly in scenarios involving high density, multiple points of interest, and communication cost constraints.
Thereafter, we propose our concept, which utilizes discretized spatial cells to represent complex-shaped geofences.
Second, to mitigate the above challenges, we propose discrete geofence design problems that offer greater flexibility than circular geofences.
Finally, we demonstrate its potential applications in mobile notifications, regional information delivery, and urban navigation design.
Notably, by introducing distance-weighted coverage, we mitigate the problem of selecting irrelevant distant cells, revealing that meaningful notification areas can be automatically configured in practical terms.
This research demonstrates the feasibility of geofence design problems, which transcends conventional circular regions, and serves as the foundational design paradigm for future practical applications.

\section*{Data availability}
The data used and analyzed during the current study are available from the corresponding author upon reasonable request.

\bibliography{reference}

\section*{Author information}
\subsection*{Contributions}
K.O. proposed a basic concept.
K.O., A.O., and H.Y. discussed the conceptualization.
K.O., A.O., and T.M. investigated formulations and methodology.
K.O. conducted the numerical experiments.
All authors analyzed the experimental results and reviewed the manuscript. 

\subsection*{Corresponding author}
Correspondence to Keisuke Otaki.

\section*{Ethics declarations}
\subsection*{Competing interests}
The authors declare no competing interests.

\end{document}